\documentclass[twocolumn]{aastex701}

\usepackage{amsmath}
\usepackage{tensor}

\newcommand{\ts}{\tilde{s}}
\newcommand{\ttheta}{\tilde{\theta}}
\newcommand{\tx}{\tilde{x}}
\newcommand{\tA}{\tilde{A}}
\newcommand{\tf}{\tilde{f}}

\newcommand{\pixsize}{\varepsilon}
\newcommand{\pixarea}{{\varepsilon^2}}
\newcommand{\pixareai}{{\varepsilon_i^2}}
\newcommand{\phibar}{\overline{\phi}}

\newcommand{\SNR}{\mathrm{SNR}}

\newcommand{\SNRmax}{\SNR_\mathrm{max}}
\newcommand{\SNRobs}{\SNR_\mathrm{obs}}

\newcommand{\erf}{\mathop{\mathrm{erf}}}
\newcommand{\qzero}{q^{(0)}}
\newcommand{\qtwo}{q^{(2)}}

\newcommand{\Dsmax}{\Delta s^2_\mathrm{max}}
\newcommand{\trialdensity}{\rho_\mathrm{trial}}

\newcommand{\Hten}{H_{10}}

\newcommand{\au}{\mathrm{au}}

\newcommand{\dsun}{d_\odot}

\DeclareMathOperator{\Diag}{Diag}

\newcommand{\skyarea}{A_\mathrm{sky}}
\newcommand{\vmax}{v_\mathrm{max}}

\shorttitle{Multi-year digital tracking}
\shortauthors{Geringer-Sameth, Golovich, and Iwabuchi}

\begin{document}

\title{Multi-year stacking searches for solar system bodies}

\author{Alex Geringer-Sameth}
\affiliation{Physics Division, Lawrence Livermore National Laboratory, 7000 East Avenue, Livermore CA, 94550, USA}
\email[show]{geringersame1@llnl.gov}

\author{Nathan Golovich}
\affiliation{Physics Division, Lawrence Livermore National Laboratory, 7000 East Avenue, Livermore CA, 94550, USA}
\email[show]{golovich1@llnl.gov}

\author{Keita Iwabuchi}
\affiliation{Center for Applied Scientific Computing, Lawrence Livermore National Laboratory, 7000 East Avenue, Livermore CA, 94550, USA}
\email[show]{iwabuchi1@llnl.gov}

\begin{abstract}
Digital tracking detects faint solar system bodies by stacking many images along hypothesized orbits, revealing objects that are undetectable in every individual exposure. Previous searches have been restricted to small areas and short time baselines. We present a general framework to quantify both sensitivity and computational requirements for digital tracking of nonlinear motion across the full sky over multi-year baselines. We start from matched-filter stacking and derive how signal-to-noise ratio (SNR) degrades with trial orbit mismatch, which leads to a metric tensor on orbital parameter space. The metric defines local Euclidean coordinates in which SNR loss is isotropic, and a covariant density that specifies the exact number of trial orbits needed for a chosen SNR tolerance. We validate the approach with Zwicky Transient Facility (ZTF) data, recovering known objects in blind searches that stack thousands of images over six years along billions of trial orbits. We quantify ZTF's sensitivity to populations beyond 5~au and show that stacking reaches most of the remaining Planet~9 parameter space. The computational demands of all-sky, multi-year tracking are extreme, but we demonstrate that time segmentation and image blurring greatly reduce orbit density at modest sensitivity cost. Stacking effectively boosts medium-aperture surveys to the Rubin Observatory single-exposure depth across the northern sky. Digital tracking in dense Rubin observations of a 10~sq.~deg field is tractable and could detect trans-Neptunian objects to 27th magnitude in a single night, with deep drilling fields reaching fainter still.
\end{abstract}

\keywords{}

\section{Introduction\label{sec:intro}}

New solar system bodies are discovered primarily in a two-step procedure: source detection in individual images followed by the linking of position changes from night to night in a self-consistent, physical orbit. However, this approach is limited to the brightest objects, those that lie above the single-exposure detection threshold. At the same time, solar system bodies adhere to an iron fact of astronomy: for every object large enough and near enough to be detected in current imaging, there are vast numbers that lie below threshold \citep[e.g.][]{2014ApJ...782..100F,2018Icar..312..181G,2018Icar..303..181J}.

These faint bodies still contribute photons to the camera, and their presence is subtly imprinted on the image data. They can be recovered by stacking images over multiple epochs, centering each image on the predicted position of the moving object, and repeating the procedure for many trial orbits. This approach is known as digital tracking, synthetic tracking, or track-before-detect in the solar system community, but is equivalent to methods used throughout astronomy (image coaddition, matched filtering). Digital tracking on deep stacks of images offers the opportunity to make striking detections of ``invisible objects'' that are well below the noise floor in every individual image.

The history of digital tracking goes back at least three decades. It has most often been used in searches for trans-Neptunian and Kuiper Belt objects (TNOs and KBOs) in the outer solar system \citep{1995ApJ...455..342C, 1998AJ....116.2042G, 1999AJ....118.1411C, 2001ApJ...549L.241A, 2004AJ....128.1364B, 2004Natur.430..865H, 2008Icar..195..827F, 2009ApJ...696...91F, 2019AJ....157..119W, 2020PSJ.....1...81R, 2024PSJ.....5..227F, 2024AJ....167..136S}, but also for the main belt \citep{2015AJ....150..125H,2019AJ....158..232H,2020PASP..132f4502Z,2023MNRAS.521.4568B} as well as for near-Earth objects (NEOs) and other inner solar system populations \citep{2014ApJ...782....1S, 2014ApJ...792...60Z, 2018AJ....156...65Z, 2018ChA&A..42..433W, 2021AJ....161..282L}. \citet{2012MNRAS.425..862G} explore the related idea of detecting a sub-threshold population of moving sources without identifying any particular object.

Stacking increases sensitivity since detector noise averages out over multiple observations while the signal from the source object adds coherently. This generally leads to the minimum detectable source flux scaling as $F_\mathrm{min} \propto 1/\sqrt{N}$, where $N$ is the number of images in the stack. From this scaling, it would seem that digital tracking is particularly well-suited to large-area sky surveys that operate over many years. Such surveys will capture a large fraction of the solar system's minor planets in thousands of individual exposures each.

However, digital tracking has so far been applied over relatively short time periods and narrow fields of view. The reason is that the computations are greatly simplified when the possible trajectories through a stack of images are linear and parallel over the duration of the observations. This limits the time baseline to around an hour for NEOs, one night for the main belt, and several days for KBOs \citep{2015AJ....150..125H}. Beyond these limits, curvature due to parallax and the ellipticity of the orbit cannot be ignored, the mapping from orbital parameters to pixel coordinates becomes expensive, and the stacking can no longer be done with an efficient shift-and-add procedure over whole images.

In this work, we consider the general digital tracking problem, in which solar system bodies must be modeled as moving on physical, nonlinear orbits across the full sky. The aim is to quantify (1) the sensitivity of optimal stacking for an arbitrary observing strategy and (2) the scale of the computational challenge of searching the parameter space of possible orbits.

We do this by first deriving how the detection significance varies as the trial orbit used to stack images departs from a moving object's true orbit. This result is used to define a metric tensor on the space of orbital parameters, which quantifies how ``distinguishable'' a pair of nearby orbits are in a given survey. This helps define the proper spacing of trial orbits: it is wasteful to search along on multiple orbits that are indistinguishable, while a search that does not sample orbits densely enough will miss detectable objects. The metric can be used to construct a local coordinate system for each region of parameter space. In local coordinates, a unit-spaced lattice corresponds to a dense set of ``barely distinguishable'' trial orbits. The metric also leads to a natural notion of trial orbit density. This density, integrated over a region of orbital parameter space, gives the number of trial orbits that must be searched to fully cover the region, i.e. to detect all moving objects to within some tolerance of their maximum possible detection significance.

The spacing of trial orbits is a decision that must be made in every digital tracking search, whether in the linear regime or not. In previous work, these spacings, for the most part, have been derived using heuristic arguments, e.g. that the on-sky tracking error be about a PSF width over the time span of the observations. Our theory gives a result similar to the heuristic methods in the linear regime and also applies to arbitrarily complicated orbital motion.

A few previous studies have considered nonlinear digital tracking. \citet{2004AJ....128.1364B} notably carried out a nonlinear TNO search using the Hubble Space Telescope in stacks of $N=55$ images taken over approximately 24~hours. The time span and observing geometry were such that the 4-d parameter space of position and velocity used in linear tracking had to be augmented with a distance dimension. A heuristic argument similar to that used for the linear regime was employed to determine the gridding in distance.

\citet{2010PASP..122..549P} develop a technique to construct sets of trial shift vectors used for image stacking. They first derive a scaling relation on the number of required trial orbits in the linear regime by considering, as we do, a tolerance on reduction of signal-to-noise ratio (SNR). For nonlinear motion, their constructive method works by generating a large library of random orbits and then prunes these to an optimum minimal set that maintains a maximum tracking error in every image. In comparison, our local coordinates can be used to construct a grid of orbits directly, while alternatively, our metric density can be used to ``fill up'' a region with a suitable number of trial orbits, similar to the Parker and Kavelaars method but without the pruning step.

A third work on nonlinear digital tracking is \citet{2018PASP..130g4504G}. They present a method for constructing shift vectors used to stack space-based observations over short time intervals, where the nonlinear motion is due entirely to the spacecraft orbiting Earth. They construct a ``zero-motion template'' corresponding to a stationary asteroid at a distance of 1~au and then apply a series of linear transformations to generate other possible shifts to stack on. The procedure is highly efficient for this specific use case. Our method, by contrast, is agnostic to the parameterization and behavior of the motion model and could accommodate an Earth-orbiting observer as a particular case.

The mathematics of our metric resembles the linear covariance approximation in nonlinear least squares fitting of asteroid orbits to observations \citep[e.g.][]{1993Icar..104..255M}. In our case, we start from the expected SNR of a matched filter in image space rather than the $\chi^2$ of sky-coordinate residuals, although the practical difference is minor. Our transformation to local coordinates by diagonalizing the metric has come up previously in mapping out the nonlinear sky-plane uncertainty for known asteroids, the so-called Line (or Volume) of Variations \citep{1996MNRAS.280.1235M, 1999Icar..137..269M, 2005A&A...431..729M, 2006MNRAS.368..809M}. The two applications have many parallels, but we are concerned with efficiently covering a large parameter volume to discover new solar system bodies in a blind search rather than characterizing known ones.

The remainder of the paper is structured as follows. Section~\ref{sec:metricintro} motivates the metric by using a heuristic notion of distance between two orbits in a stack of images. In Sec.~\ref{sec:metricfromSNR} we rederive the metric as measuring the relative reduction in detection significance caused by using the incorrect orbit in the matched filter search. The analytic formulas are verified by comparing them with simulations and Zwicky Transient Facility (ZTF) images of the minor planet Sedna (Sec.~\ref{sec:sedna}). In Sec.~\ref{sec:normalcoords} we construct a local coordinate system for the space of possible orbits, and in Sec.~\ref{sec:density} introduce the notion of density of trial orbits, again validating the theory in experiments with ZTF. Next, in Sec.~\ref{sec:linearmotion} we apply the methods to the special case of linear motion, suitable for relatively short time spans and limited sky areas. In Sec.~\ref{sec:ZTF}, we quantify the reach and difficulty of fully coherent digital tracking with six years of ZTF data, covering about ${30{,}000~\deg^2}$ of the northern sky \citep{2019PASP..131a8002B}. Finally, in Sec.~\ref{sec:discussion} we discuss multi-year digital tracking in relation to the Rubin Observatory's Legacy Survey of Space and Time (LSST) as well as tradeoffs between computational cost and sensitivity. In the appendix, we give a careful derivation of the expected detection significance for general optimal image stacking, which is used in Sec.~\ref{sec:metricfromSNR}.

\section{A metric in orbital parameter space\label{sec:metricintro}}

\subsection{Motivation}
Digital tracking can be thought of as a trial-and-error approach to the discovery of moving objects. In each trial, an orbit is suggested and then the imaging data is analyzed to see whether there is evidence of a solar system body on that orbit. This is an example of forward modeling in that we start from the physical parameters describing the orbit and then predict what the signal looks like in data space, i.e. determining which survey images and pixels within those images contain the object. The detection significance is calculated by centering the images on the predicted location of the moving object, stacking them (with weights), and looking for excess flux at the center of the stacked image.

The true orbital parameters of the body are not known beforehand, and the search proceeds by enumerating a long list of trial orbits, testing each one for the presence of a signal. The question is how to generate a suitable set of trial orbits.

A useful concept here is the density of trial orbits in the parameter space. If the density is too high, two trial orbits may be so similar that they intersect the same survey images and the same pixels within those images. In this case, the detection significance is highly correlated between these orbits. There will be either a detection on both orbits or on neither one. The opposite problem are voids, regions of orbital parameter space far from any trial orbit. If there is an actual solar system body in such a region, it will go undetected.

One approach is to construct a lattice of points in parameter space that have the correct spacing. This can be done by developing a distance measure in the space of possible orbits that quantifies how ``distinguishable'' two orbits are in the imaging data. The optimal spacing of the lattice is such that two adjacent orbits are barely distinguishable.

\subsection{Heuristic derivation of the metric in image space\label{sec:heuristicmetric_imagecoords}}

By what measure should we determine whether two orbits are distinguishable? The answer is given by the angular resolution or point spread function (PSF) of the imaging. The PSF can provide a metric in image space which we can then be pulled back to the parameter space of orbits.

Consider a trial orbit that intersects $N$ survey images at pixel coordinates $(x_i, y_i)$ for $i=1,\ldots,N$. In other words, a real object on that orbit would be present at location $(x_i, y_i)$ in image $i$. Now imagine a slightly different orbit which shifts the object's coordinates in each image by a small amount $(\Delta x_i, \Delta y_i)$. Denoting the width of the PSF in image $i$ by $b_i$, a reasonable metric for the distinguishability of the two orbits is
\begin{align}
\Delta s^2  &= \frac{1}{N} \sum\limits_{i=1}^N \frac{\Delta x_i^2 + \Delta y_i^2}{b_i^2}, \label{eqn:metric_imagecoords_xy}
\end{align}
with the criterion for distinguishability being
\begin{equation}
\Delta s^2 \gtrsim 1. \label{eqn:heuristicdistinguishabilitycriterion}
\end{equation}

If $\Delta s^2 < 1/N$, it is guaranteed that the difference between the orbits is less than a PSF width in every image. If there were an actual object on one of the orbits, stacking images along either one will give nearly identical detection significances. As $\Delta s^2$ approaches 1, the position shift is on average about 1 PSF width in each image. In this case, an object on one of the orbits will not contribute much flux at the location of the other orbit, and the detection significance for the two orbits will begin to decorrelate.

Setting the distinguishability threshold at $\Delta s^2 \approx 1$ instead of $\Delta s^2 \approx 1/N$ assumes that the shifts in different images are of similar magnitude. This is reasonable as there are generally many more image intersections than the number of degrees of freedom of possible orbits. It would be unusual to adjust one of the few orbital degrees of freedom and have the orbit's location remain unchanged in nearly every image except for a few in which it shifts by a large amount.

\subsection{Pulling back to the space of orbits \label{sec:heuristicmetric}}

The metric defined in Eq.~\ref{eqn:metric_imagecoords_xy} induces a metric on the space of possible orbits\footnote{In this paper we will refer to the space of possible orbits as the parameter space, orbital space, or phase space.}. To see this, it will help to rewrite the metric in the language of differential geometry.

For a given trial orbit, let $x^i$ be the image coordinates of an imaginary object on that orbit. The index $i$ is a multi-index which runs over both the survey images containing the object and over the two dimensions ($x$ and $y$) in each image.

In this notation, the metric from Eq.~\ref{eqn:metric_imagecoords_xy} takes the form
\begin{align}
\Delta s^2 &= \eta_{ij} \, \Delta x^i \Delta x^j, \label{eqn:metric_imagecoords}
\end{align}
where
\begin{align}
\eta_{ij} &= \frac{1}{N}\frac{1}{b_i^2} \,\delta_{ij} = \frac{1}{N}\mathrm{Diag}\left(\frac{1}{b_i^2}\right). \label{eqn:heuristic_eta_def}
\end{align}

Equation~\ref{eqn:metric_imagecoords} and subsequent equations use the Einstein summation convention, where a repeated upper and lower index implies summation over all values the index can take (i.e. $\eta_{ij} \, \Delta x^i \Delta x^j$ means $\sum_i \sum_j \eta_{ij} \, \Delta x^i \Delta x^j$). In Eq.~\ref{eqn:heuristic_eta_def}, $\delta_{ij}=1$ when $i=j$ and 0 otherwise. It is straightforward to generalize Eq.~\ref{eqn:heuristic_eta_def} for an elliptical PSF, but below we only consider isotropic PSFs with a single $b_i$ for the both the $x$ and $y$ directions in each image.

Now we consider the space of possible orbits. In linear digital tracking, phase space is four-dimensional: an orbit is specified by an initial position and angular velocity in the plane of the sky. In the general case of moving bodies in the solar system, the parameter space is six-dimensional. For concreteness, we use the Keplerian elements (semi-major axis, eccentricity, inclination, argument of perihelion, longitude of the ascending node, and mean anomaly at a given epoch), but the formalism equally applies to any other element sets, e.g. equinoctal elements, the moving object's 3-d position and velocity at a fixed moment in time, or the angular position plus parallax parameterization of \citet{2000AJ....120.3323B}. As a geometric object, the metric we will derive is fundamentally independent of the choice of phase space coordinates.

Coordinates in orbital parameter space will be grouped into the vector $\theta^\mu$, where $\mu=1,2,\ldots,6$ (e.g. for Kepler elements $\theta^1$ is semi-major axis, $\theta^2$ is eccentricity, etc).

The forward model from orbital parameters to image space has two steps. First, the orbital parameters are used to determine where in the sky the solar system body was during each exposure of the survey. This yields a list of intersections, i.e. survey images that should contain an object on that orbit. Next, for each intersection there is a mapping from sky coordinates to image coordinates (also called pixel coordinates). The second step is done with either a basic tangent plane projection or with the FITS World Coordinate System \citep[WCS,][]{2002A&A...395.1077C} solution for the image.

The forward model is a mapping from $\theta^\mu$ to $x^i$. In other words, a given set of orbital elements maps to a set of pixel coordinates in our survey images. The key to the metric is the Jacobian of this mapping,
\begin{align}
J^i_\mu (\theta)\equiv \frac{\partial x^i}{\partial \theta^\mu}. \label{eqn:jacobiandef}
\end{align}
If the orbit $\theta$ under consideration intersects $N$ survey images, the Jacobian $J^i_\mu$ can be thought of as a $2N \times 6$ matrix $J$ (the 2 coming from the $x$ and $y$ coordinates for each image). The Jacobian is straightforward to compute numerically, either through finite differencing (what we do) or automatic differentiation (which requires coding the forward model using a framework like PyTorch or JAX).

The Jacobian describes how a small change in a solar system body's orbital elements $\Delta \theta^\mu$ maps to a small change in image coordinates $\Delta x^i$,
\begin{equation}
\Delta x^i = J^i_\mu \Delta \theta^\mu. \label{eqn:linearmap_orbitalparams_imageparams}
\end{equation}
Therefore, the distinguishability metric between the two orbits $\theta^\mu$ and $\theta^\mu + \Delta \theta^\mu$ is
\begin{align}
\Delta s^2 &= \eta_{ij}\Delta x^i \Delta x^j \nonumber \\
&= g_{\mu\nu}  \Delta \theta^\mu \Delta \theta^\nu, \label{eqn:metric_orbitspace}
\end{align}
where $g_{\mu\nu}$ is defined by
\begin{equation}
g_{\mu\nu} = J^i_\mu \eta_{ij} J^j_\nu. \label{eqn:metrictensor_orbitspace}
\end{equation}

The object $g_{\mu\nu}$ is a metric tensor on orbital parameter space. It can be thought of as the $6 \times 6$ matrix $g$,
\begin{equation}
g = J^T \eta J, \label{eqn:metric_matrixdef}
\end{equation}
where $\eta$ is the diagonal matrix defined in Eq.~\ref{eqn:heuristic_eta_def}.

Importantly, the metric is not constant but is a function of location within orbital parameter space. The space of possible orbits has become a Riemannian manifold with curvature. This is an example of information geometry \citep[e.g.][]{amari1985differential}: the sum of squared offsets in Eq.~\ref{eqn:metric_imagecoords_xy} is essentially a chi-square or negative log-likelihood function and the metric $g_{\mu\nu}$ is the Fisher information matrix of second derivatives of this log-likelihood with respect to the model parameters.

\begin{figure}[b!]
\begin{center}
\includegraphics{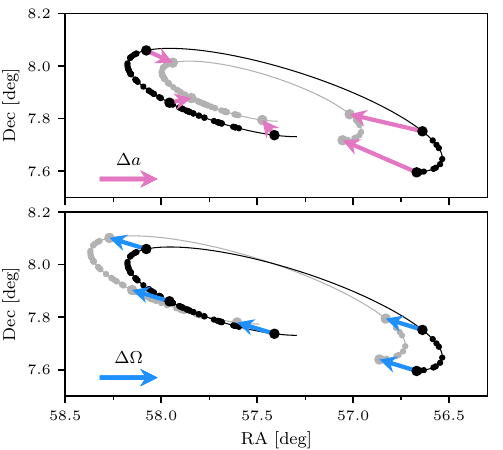}
\end{center}
\caption{\label{fig:orbitonsky_deltaelems} Illustration of the change in sky coordinates of an orbit under changes $\Delta\theta^\mu$ in the orbital parameters. Black curves show the path of Sedna during 2019. Points mark the intersections with ZTF, i.e. times when the object would be present in a survey image. The gray curves show how the orbital path shifts under a change in semi-major axis $a$ (top) or longitude of the ascending node $\Omega$ (bottom). The arrows illustrate the Jacobian of the transformation. The changes in orbital parameters are exaggerated for illustration. In practice the shifts used in digital tracking will be at the arcsecond level.}
\end{figure}
Figure~\ref{fig:orbitonsky_deltaelems} illustrates the transformation from orbital parameters to sky coordinates as seen by the Zwicky Transient Facility survey \citep[ZTF;][]{2019PASP..131a8002B,2019PASP..131f8003B}. In each panel, the black curve shows the path of the dwarf planet Sedna during 2019. The dots mark the times that Sedna is present in a ZTF survey image, i.e. each dot denotes an intersection with the survey at position $x^i$. The gray curves and points show how the $x^i$'s would change under a change $\Delta\theta^\mu$ in two of the orbital parameters (upper: semi-major axis, lower: longitude of the ascending node). The arrows illustrate the Jacobian of the transformation $\partial x^i/\partial \theta^\mu$. The figure shows very large shifts in the orbital parameters for visualization. In practice, we will be concerned with changes in orbital elements $\Delta \theta^\mu$ that change the position on the sky by arcseconds.

\section{The metric derived from expected detection significance\label{sec:metricfromSNR}}
The previous section provided an intuitive motivation for the image space metric in Eq.~\ref{eqn:metric_imagecoords_xy}. The shift of the orbit in image space is only a proxy for what we really care about in digital tracking: how detection sensitivity falls off as the trial orbit deviates from an object's true orbit. Here, we derive a metric based directly on expected detection significance. This is the geometric approach introduced by \citet{1996PhRvD..53.6749O} to construct template banks for gravitational wave searches, building on the work of \citet{1991PhRvD..44.3819S} and \citet{1994PhRvD..49.1707D}.

\subsection{Optimal image stacking \label{sec:optimalimagestacking}}
The optimal linear method of combining observations to detect a faint source is matched filtering \citep[also called pixel weighting, Weiner filtering, or linear filtering; see, e.g.][]{1992ApJ...398..169R, Press2007}. In digital tracking, after a trial orbit predicts the location of a moving object in each image, the nearby pixels in all images are combined into a scalar test statistic via a weighted sum. The expected detection significance is \citep[see][Sec.~IV]{2015PhRvD..91h3535G}
\begin{align}
\SNR = \frac{\sum\limits_i w_i s_i}{\sqrt{\sum\limits_i w_i^2 \sigma_i^2}} \label{eqn:SNR_weights},
\end{align}
where $w_i$ is the weight assigned to pixel $i$, $s_i$ is the expected flux in pixel $i$ due to the object, and $\sigma_i^2$ is the variance in the pixel due to all background processes (this assumes that the noise in different pixels is uncorrelated). The sums are over all pixels in all the images the trial orbit intersects.

As a measure of detection significance, the SNR tells us how many standard deviations above background level the source will be detected at. It is a dimensionless quantity, i.e., $\SNR=5$ is a ``5 sigma'' detection.

The optimal pixel weights for source discovery are \citep[][Eq.~16]{2015PhRvD..91h3535G},
\begin{equation}
w_i = \frac{s_i}{\sigma_i^2}. \label{eqn:optweights}
\end{equation}

Note that the expected significance in Eq.~\ref{eqn:SNR_weights} is unchanged if all weights are scaled by a constant factor. This is important since it implies the absolute magnitude of a solar system body need not be known in advance. One can choose any absolute magnitude when calculating $s_i$ for the weights in Eq.~\ref{eqn:optweights} and the search remains optimal.

\subsection{Analytic form for expected significance}

In practice, it is impossible to carry out optimal weighting since we do not know the true orbit of a solar system body before we search for it. The pixel weights (Eq.~\ref{eqn:optweights}) are instead calculated using a trial orbit. The relevant question is how the sensitivity decreases as a function of the difference between the true orbit and the trial orbit. To answer this, we start from a closed-form expression for the SNR (Eq.~\ref{eqn:SNR_weights}) in terms of the image-space offsets $\Delta x^i$ between the trial and true orbits.

The theory is worked out in Appendix~\ref{sec:app_expectedSNR}; we summarize the results here. If a point source's true position in image $i$ is $x^i$, but the images are stacked using trial positions $\tx^i = x^i + \Delta x^i$ ($\Delta x^i$ being a 2-d vector for each image), the expected detection significance is
\begin{align}
\SNR(\Delta x) &= \SNRmax \sum\limits_{i=1}^N c_i \frac{\phibar_{\sqrt{2}b_i}(\Delta x^i)}{\phibar_{\sqrt{2}b_i}(0)}. \label{eqn:SNRexact}
\end{align}
The sum is over the images $i$ that the orbits intersect ($\Delta x^i$ will always be small enough that the trial and true positions are in the same images). The PSF in image $i$ is modeled as a 2-d Gaussian $\phi_b$ (Eq.~\ref{eqn:2dgaussianpdf}) with standard deviation $b_i$, and $\phibar_b$ is the Gaussian density averaged over an image pixel (defined in Eq.~\ref{eqn:avggaussian}; approximately equal to $\phi_b$ for small pixel sizes).

The $c_i$ in Eq.~\ref{eqn:SNRexact} are a set of image weights that sum to $1$ and $\SNRmax$ is the expected significance when the trial positions are equal to the true positions, i.e. $\SNRmax$ is maximum SNR that could be achieved with an infinite density of trial orbits. These are given by
\begin{equation}
\SNRmax = \dfrac{\displaystyle\sum\limits_i \dfrac{A_i^2 \qzero_i}{4\pi \sigma_i^2 (b_i/\pixsize_i)^2}}
{\displaystyle\sqrt{\sum\limits_i \dfrac{A_i^2}{4\pi \sigma_i^2 (b_i/\pixsize_i)^2}}} \label{eqn:SNRmaxexact}
\end{equation}
and
\begin{equation}
c_i = \frac{c'_i}{\sum_i c'_i}, \quad \text{where}\quad c'_i = \frac{\tA_i^2 \qzero_i}{\sigma_i^2 (b_i/\pixsize_i)^2}. \label{eqn:SNRc_exact}
\end{equation}
In these equations, $A_i$ is the total flux of the source in image $i$ (with the same units as $\sigma_i$, e.g. ADU), while $\tA_i$ is the predicted flux at the trial positions (only relative flux differences between images matter for $\tA_i$, not their absolute scaling). As above, $\sigma_i^2$ is the pixel noise variance (now assumed the same for all pixels near the source in image $i$), $\pixsize_i$ is the pixel scale (so that $b_i/\pixsize_i$ is the PSF size in pixel units), and $\qzero_i$ (shorthand for $\qzero_{\pixsize_i/\sqrt{2}b_i}$, defined in Eq.~\ref{eqn:finitepixcorrection0def}) is a finite pixel size correction factor that converges to 1 for small pixels (i.e. when $\pixsize_i \ll b_i$). The above equations can be found in Appendix~\ref{sec:app_expectedSNR_finite}.

As a function of $i$, the sequence $A_i$ is the solar system body's light curve. The light curve is a function of the Sun-body-observer geometry as well as intrinsic properties of the body, traditionally parameterized by an absolute magnitude $H$ and slope parameter $G$ in the ``H,~G system'' \citep{1989aste.conf..524B}. Absolute magnitude does not need to be searched over as it is an overall scaling in $\tA_i$. It would be straightforward to augment orbital parameter space to include $G$, but for efficiency we fix it at the conventional value $G=0.15$, leading to a negligible loss in sensitivity except in very peculiar circumstances involving the opposition surge effect. In what follows, the difference between the trial and true orbits will always be so small that the trial light curve is proportional to the true light curve ($\tA_i \propto A_i$). The constant of proportionality depends on the object's (unknown) absolute magnitude.

\subsection{From expected significance to metric \label{sec:exactmetric}}
Equation~\ref{eqn:SNRexact} describes how the expected detection significance falls off as the location of the trial object deviates from the true location, i.e. as a function of $\Delta x^i$. We now proceed as in Sec.~\ref{sec:heuristicmetric} to relate this falloff to the separation of the two orbits in orbital parameter space, $\Delta\theta^\mu = \ttheta^\mu - \theta^\mu$, where $\ttheta$ and $\theta$ denote the trial and true orbits. 

When the trial and true orbits are similar, the separations $\Delta x^i$ in the individual images are small. Expanding Eq.~\ref{eqn:SNRexact} to lowest order in $\Delta x^i$ we have
\begin{multline}
\SNR(\Delta x)= \SNRmax \left(1 - \sum\limits_i \frac{1}{4} c_i  \qtwo_i \left\lvert \frac{\Delta x^i}{b_i}\right\rvert^2  \right.\\
 + \left. \mathcal{O}\left\lvert\frac{\Delta x^i}{b_i}\right\rvert^4  \right), \label{eqn:SNRtaylor}
\end{multline}
where we used the Taylor series from Eq.~\ref{eqn:avggaussianseries}, the finite pixel size correction factor $\qtwo_i=\qtwo_{\pixsize_i/\sqrt{2}b_i}$ defined in Eq.~\ref{eqn:finitepixcorrection2def}, and the fact that $c_i$ sum to 1 (Eq.~\ref{eqn:SNRc_exact}).

There is no linear $\Delta x^i$ term because the test statistic (i.e. set of optimal weights) is derived from a minimization procedure \citep{2015PhRvD..91h3535G}. This means that first order changes in the weights (i.e. changes in the trial orbit) lead to second-order changes in detection significance. This is a general feature of optimal linear (Weiner) estimators \citep[e.g.][]{1992ApJ...398..169R} and is what allows us to construct a metric, which must be quadratic in the orbital parameters $\Delta\theta^\mu$.

Next we rewrite the quadratic term in tensor notation as in Eqs.~\ref{eqn:metric_imagecoords} and~\ref{eqn:heuristic_eta_def} (dropping the $\mathcal{O}\lvert \Delta x/b\rvert^4$ terms from now on),
\begin{align}
\SNR(\Delta x) &\approx \SNRmax\left(1 - \eta_{ij} \Delta x^i \Delta x^j \right) \nonumber \\
&= \SNRmax \left(1 - \Delta s^2\right), \label{eqn:SNRmetric}
\end{align}
where
\begin{equation}
\eta_{ij} = \left(\frac{1}{4} c_i \, \qtwo_i \frac{1}{b_i^2}\right)\, \delta_{ij}. \label{eqn:eta_def}
\end{equation}
As in Sec.~\ref{sec:heuristicmetric}, the indices $i$ and $j$ are multi-indices that index both images and the two dimensions within each image (though $c_i$, $\qtwo_i$, and $b_i$ are equal for the $x$ and $y$ directions within an image).

From Eq.~\ref{eqn:SNRmetric} we see that metric distance
\begin{equation}
\Delta s^2 = \eta_{ij} \Delta x^i \Delta x^j
\end{equation}
represents the fractional loss in expected detection significance when using a trial orbit that deviates from the true orbit by $\Delta x^i$ in each image.

The new metric is very similar to the heuristic one from Sec.~\ref{sec:heuristicmetric_imagecoords}. In Eq.~\ref{eqn:metric_imagecoords_xy}, each image intersection contributes to the metric with an effective weight of $1/N$. In the new metric, the contribution of image $i$ is $c_i$ (Eq.~\ref{eqn:SNRc_exact}), which captures the relative importance of the image, taking into account the noise level, light curve, and sharpness of the PSF in that image. If all intersections are of similar importance, then the $c_i$'s will all be approximately $1/N$.

We also have the same criterion (Eq.~\ref{eqn:heuristicdistinguishabilitycriterion}) for two orbits to be independent or distinguishable. From Eq.~\ref{eqn:SNRmetric} we see that when $\Delta s^2 < 1$ the expected significance is close to its maximum but drops quickly once $\Delta s^2 \gtrsim 1$. Thus the condition $\Delta s^2=1$ sets the scale at which we expect the trial positions to become suboptimal for detecting a source. 

Of course, $\SNR$ does not actually go negative when $\Delta s^2 $ reaches 1 as Eq.~\ref{eqn:SNRmetric} seems to imply. Instead, Eq.~\ref{eqn:SNRexact} (or, more transparently, Eq.~\ref{eqn:SNRsmallpix_compact}) shows that the falloff will be roughly squared exponential. Therefore, the better target for quadratic approximation is probably $\log \left[\SNR(\Delta x)\right]$ rather than $\SNR(\Delta x)$. From Eq.~\ref{eqn:SNRtaylor} or Eq.~\ref{eqn:SNRmetric} one sees that
\begin{equation}
\log\SNR(\Delta x) = \log\SNRmax - \Delta s^2 + \mathcal{O}\left\lvert\frac{\Delta x^i}{b_i}\right\rvert^4,
\end{equation}
so that
\begin{equation}
\SNR(\Delta x) \approx \SNRmax \, \exp\left(-\Delta s^2\right). \label{eqn:SNRmetricexpform}
\end{equation}
We can use this exponential form to predict that the detection significance drops to 50\% of its maximum when $\Delta s^2 \approx 0.7$ and to 10\% when $\Delta s^2 \approx 2.3$. Jensen's inequality applied to Eq.~\ref{eqn:SNRsmallpix_compact} shows that this approximation is conservative: the actual expected detection significance declines more gradually than Eq.~\ref{eqn:SNRmetricexpform} predicts. 

Finally, we retrace our steps from Sec.~\ref{sec:heuristicmetric} to transfer the metric to orbital parameter space. For small deviations of the orbital elements, we have $\Delta x^i = J^i_\mu \Delta\theta^\mu$ (Eqs.~\ref{eqn:jacobiandef} and~\ref{eqn:linearmap_orbitalparams_imageparams}), which leads to
\begin{equation}
\Delta s^2 = g_{\mu\nu} \Delta \theta^\mu \Delta\theta^\nu, \label{eqn:metricdistdef}
\end{equation}
where
\begin{equation}
g_{\mu\nu}= J^i_\mu \eta_{ij} J^j_\nu, \label{eqn:metricdef}
\end{equation}
with $\eta_{ij}$ now given by Eq.~\ref{eqn:eta_def}. The behavior of the expected significance for small orbital separations $\Delta \theta$ is given by Eq.~\ref{eqn:SNRmetricexpform}.

\subsection{Application to Sedna\label{sec:sedna}}
We demonstrate these results for the minor planet Sedna as observed by ZTF. We query Sedna's orbital elements from NASA/JPL Horizons\footnote{Data retrieved 2025-Mar-14.} \citep{NASAJPLHorizons,1996DPS....28.2504G}, propagate the orbit to the observation times of all ZTF $r$ and $g$-band difference images, and identify those that should contain Sedna. There are 971 intersections, excluding those where Sedna was in a masked part of the image. For each image we extract the PSF width and zero-point magnitude from the FITS header, estimate the pixel noise level, and calculate Sedna's predicted flux in the image (see Sec.~\ref{sec:dataprep} for details). These are used to form the pixel weights (Eq.~\ref{eqn:weight_pixcenter}) and evaluate the observed matched filter detection significance $\SNRobs$. The stacked detection significance for Sedna in ZTF is $\SNRobs = 127$, while its median $\SNR$ in individual images is 3.

Next, for each intersection we generate an ideal, zero-noise version of the difference image, where the value in each pixel is the Gaussian PSF integrated over the pixel and scaled by the predicted flux in the image. Since the predicted fluxes were calculated with an arbitrary overall scaling (arbitrary absolute magnitude $H$), the ideal images are scaled by an overall factor such that the analytic $\SNRmax$ from Eq.~\ref{eqn:SNRmaxexact} is the same as $\SNRobs$. This scaling factor corresponds to an absolute magnitude for Sedna of $H=1.87$, consistent\footnote{A fractional uncertainty in flux of $1/\SNRobs$ gives an uncertainty of $\Delta H \approx 0.01$. The agreement with \citet{2007AJ....133...26R} is perhaps coincidental since we use Makemake's colors rather than Sedna's when predicting image fluxes (see Sec.~\ref{sec:dataprep}).} with $1.829 \pm 0.048$ measured by \citet{2007AJ....133...26R}.

\begin{figure}
\begin{center}
\includegraphics{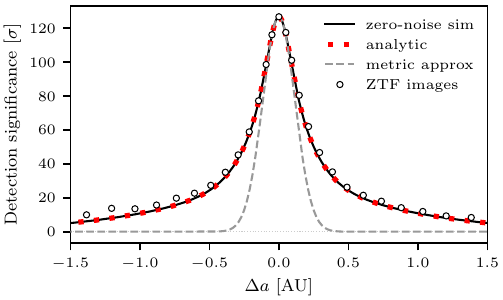}
\end{center}
\caption{\label{fig:sedna_SNR_vs_a} Detection significance in stacked images as the semi-major axis of the trial orbit deviates from Sedna's true orbit. The black curve is for ``perfect'' images simulated without noise while white points correspond to actual ZTF data. The red dotted curve is the analytic form of expected significance (Eq.~\ref{eqn:SNRexact}) and the dashed gray shows the metric approximation (Eq.~\ref{eqn:SNRmetricexpform}). The predicted significance matches the simulation and observations for all trial orbits, while the metric approximation is valid near the true orbit.}
\end{figure}
We then repeatedly perform the stacked image search for different trial orbits, varying the semi-major axis while holding the other orbital elements fixed. Figure~\ref{fig:sedna_SNR_vs_a} shows the results for both the simulated images (black curve) and the actual ZTF images (white circles). The analytic form for the expected detection significance (Eq.~\ref{eqn:SNRexact}) is plotted as the red dotted curve, showing excellent agreement with the simulated images. The significance curve for the actual ZTF data also falls off as predicted, verifying that our assumptions are safe, in particular that the PSF can be approximated as Gaussian. We note that the only free parameter that has been adjusted here is an overall scaling (the absolute magnitude of Sedna). The fact that significance in ZTF peaks at $\Delta a=0$ verifies our orbit propagation as well as the astrometric accuracy of the ZTF pipeline.

Finally, the approximate significance using the metric (Eqs.~\ref{eqn:SNRmetricexpform} and~\ref{eqn:metricdistdef}) shows good agreement with the real and simulated data near the peak (gray dashed curve). The metric tensor $g_{\mu\nu}$ (Eq.~\ref{eqn:metricdef}) is evaluated once for the true orbit. The varying trial orbit only enters through $\Delta \theta^\mu = (\Delta a, 0,0,0,0,0)$ in Eq.~\ref{eqn:metricdistdef}. The falloff in significance is notably non-Gaussian away from the peak and cannot be described globally by any single Gaussian. However, the metric serves its purpose of predicting how far the trial orbit can deviate from the true orbit before suffering a loss of significance. When the metric approximation (Eq.~\ref{eqn:SNRmetricexpform}) predicts that the significance has dropped by 10\% ($\Delta s^2 = 0.11$), the actual significance has decreased by 9.3\%. For predicted 50\% and 90\% drops ($\Delta s^2 = 0.69$ and $2.3$), the actual decreases are 37\% and 60\%.

\section{Local coordinates for orbital elements\label{sec:normalcoords}}

The metric $g_{\mu\nu}$ leads to a natural coordinate system on orbital phase space analogous to locally inertial coordinates in general relativity (GR). According to the equivalence principle, the local experience of a freely falling observer in a gravitational field is indistinguishable from flat Minkowski spacetime. Mathematically, this means that at any point in spacetime there exists a local coordinate system in which the GR metric is flat ($g_{\mu\nu}=\mathrm{Diag}(-1,1,1,1)$), and that it stays flat in the nearby vicinity ($\partial g_{\mu\nu} / \partial x^\sigma =0$), so that the coordinates represent an inertial frame in the absence of gravity \citep[][Ch.~3]{1972gcpa.book.....W}. In differential geometry such coordinates are known as (Riemann) normal coordinates. 

For digital tracking, ``locally inertial orbital elements'' make the relation between orbits and detection significance flat (i.e. Euclidean): a unit step in any direction in local coordinates changes the detection SNR by a fixed fraction. In this work, we will stop short of constructing true normal coordinates, but carry out the first step of Euclideanizing the metric at a point in phase space by a linear transformation. The next step, a quadratic transformation that zeros out the first derivatives of the metric, would be an interesting topic to explore.

One use of these local coordinates is to construct a trial orbit lattice such that every possible orbit of interest is at a maximum metric distance $\Delta s^2 < \Dsmax$ from a lattice point. This ensures that a digital tracking search completely covers a region of orbital space in the sense that every potentially detectable solar system body will be found. An optimal lattice fills the space with a minimum number of points. The metric also allows for a simple determination of the density of this optimal trial orbit lattice (Sec.~\ref{sec:density}), which quantifies the computational difficulty of the search in various regions of orbital parameter space.

\subsection{Coordinate transformation}
Normal coordinates, denoted $\xi^\mu$, are an alternate set of orbital elements that parameterize the possible orbits in the neighborhood of a particular orbit $\theta_0^\mu$. In other words, there is an invertible mapping between the two sets of coordinates, $\theta^\mu = \theta^\mu(\xi^\nu)$, where $\xi^\mu=0$ maps to $\theta^\mu =\theta_0^\mu$. In the new coordinates, the metric at the point $\xi^\mu$ has components $h_{\mu\nu}(\xi)$ given by
\begin{equation}
h_{\mu\nu}(\xi) = g_{\alpha\beta}(\theta) \frac{\partial \theta^\alpha}{\partial \xi^\mu} \frac{\partial \theta^\beta}{\partial \xi^\nu}, \label{eqn:metricchangecoords}
\end{equation}
which is just the statement that the infinitesimal metric distance between the two orbits $\xi^\mu$ and $\xi^\mu+d\xi^\mu$ is $ds^2 = h_{\mu\nu} d\xi^\mu d\xi^\nu$.

The actual transformation is built by examining Eq.~\ref{eqn:metricchangecoords} order by order in $\xi^\mu$. Specifically, the mapping is written \citep[e.g.][]{brewinnormalcoordnotes}
\begin{equation}
\theta^\mu = \theta_0^\mu +  \Lambda\indices{^\mu_\nu} \xi^\nu + \frac{1}{2} B\indices{^\mu_\nu_\rho} \xi^\nu \xi^\rho + \cdots, \label{eqn:coordtransform_powers}
\end{equation}
and the constant coefficients $\Lambda\indices{^\mu_\nu}$ and $B\indices{^\mu_\nu_\rho}$ are chosen so that $h_{\mu\nu}$ is diagonal and has first derivatives equal to zero. We only consider the  linear transformation $\Lambda\indices{^\mu_\nu}$ to keep things simple, and set all higher coefficients to zero.

\subsection{Linear transformation of the orbital elements}
The goal here is to choose the linear part of the transformation $\Lambda\indices{^\mu_\nu}$ so that at $\xi^\mu=0$ the metric takes the Euclidean form $h_{\mu\nu}(0) =\delta_{\mu\nu}$. From Eqs.~\ref{eqn:metricchangecoords} and~\ref{eqn:coordtransform_powers}, we have
\begin{equation}
h_{\mu\nu}(0) = g_{\alpha\beta}(\theta_0) \Lambda\indices{^\alpha_\mu} \Lambda\indices{^\beta_\nu}  = \delta_{\mu\nu}. \label{eqn:metric0_changecoords}
\end{equation}
Finding $\Lambda\indices{^\mu_\nu}$ means solving the matrix equation
\begin{equation}
\Lambda^T g \Lambda = I, \label{eqn:linear_euclidean_transf_matrixform}
\end{equation}
where $g$ is the metric in $\theta$-coordinates evaluated at $\theta_0$, $I$ is the identity matrix, and $\Lambda\indices{^\mu_\nu}$ is the row $\mu$ column $\nu$ entry of the square matrix $\Lambda$.

Equation~\ref{eqn:linear_euclidean_transf_matrixform} specifies $\Lambda$ only up to an overall rotation of coordinates\footnote{This is likely related to the observation by \citet{2005A&A...431..729M} that the Line of Variations and ``weak direction'' depend on the choice of orbital elements.} (if $\Lambda$ solves Eq.~\ref{eqn:linear_euclidean_transf_matrixform} so does $\Lambda O$ for any orthogonal matrix $O$). A particular solution is to let each column of $\Lambda$ be a normalized eigenvector of $g$ scaled by the inverse square root of the corresponding eigenvalue. This solution always exists (i.e. the eigenvalues are positive and eigenvectors mutually orthogonal) since the metric $g$ (Eq.~\ref{eqn:metricdef}) is positive definite (which in turn comes from the diagonal image-space metric $\eta_{\mu\nu}$ (Eq.~\ref{eqn:eta_def}) having all positive entries).

Numerically, one can diagonalize $g$ directly (which is efficient for a $6\times6$ matrix), but it is more revealing to arrive at the transformation by singular value decomposition \cite[SVD; e.g.][Ch.~2.6]{Press2007}. The metric $g$ is first factorized (see Eq.~\ref{eqn:metric_matrixdef} or~\ref{eqn:metricdef}) as
\begin{equation}
g = J^T \eta J = \left(\sqrt{\eta} J \right)^T \left(\sqrt{\eta} J \right), \nonumber
\end{equation}
where the square root of the positive diagonal matrix $\eta$ (Eq.~\ref{eqn:eta_def}) is taken elementwise. At this point $\sqrt{\eta}J$ has units, which makes matrix decomposition awkward. The values in column $\mu$ have units of {1/[units of $\theta^\mu$]} so a choice of length scale for each orbital element can be used to remove them. \citet{2005A&A...431..729M} suggest some simple scalings for different sets of orbital elements (e.g. for Kepler elements, dividing semi-major axis by $a_0$, inclination by $180^\circ$, and the other angular elements by $360^\circ$). We find it natural to use the metric itself to define a length scale $\ell^\mu$ for each coordinate,
\begin{equation}
\ell^\mu = \left(g_{\mu\mu}\right)^{-1/2}. \label{eqn:elemscales}
\end{equation}
Taking a step in an orbital element $\theta^\mu \to \theta^\mu + \ell^\mu$ and holding the others constant gives a metric change of $\Delta s^2=1$, meaning $\ell^\mu$ is a step size that is just distinguishable in the data.

Let $\ell$ be the diagonal matrix $\Diag(\ell^\mu)$. Then $\sqrt{\eta} J \ell$ is dimensionless and the SVD is performed:
\begin{equation}
\sqrt{\eta} J \ell = U D V^T. \label{eqn:SVD}
\end{equation}
The diagonal matrix $D$ ($6\times 6$) contains the square roots of the eigenvalues of $\ell g \ell$. The columns of $U$ ($2N \times 6$, for $N$ image intersections) are an orthonormal basis for the possible shifts in image-space (scaled by $\sqrt{\eta}$) corresponding to any small change in orbital elements. The columns of $V$ ($6\times 6$) are the orthonormal eigenvectors of $\ell g \ell$. A solution for $\Lambda$ is then
\begin{equation}
\Lambda = \ell V D^{-1}. \nonumber
\end{equation}
The coordinate transformation between the original orbital parameters $\theta^\mu$ and local coordinates $\xi^\mu$ is given in matrix form by
\begin{align}
\theta &= \theta_0 + \ell V D^{-1} \xi \label{eqn:xi2theta}\\
\xi &= D V^T \ell^{-1}(\theta-\theta_0), \label{eqn:theta2xi}
\end{align}
where $\theta$ and $\xi$ are vectors with components $\theta^\mu$ and $\xi^\mu$.

The linear transformation is already quite useful on its own, even without the higher-order terms in Eq.~\ref{eqn:coordtransform_powers}. The dimensionless {$\xi$-coordinates} make it trivial to calculate the drop in detection significance as a trial orbit departs from a true orbit at $\theta_0$. The fractional reduction in significance is just the squared Euclidean distance from the origin to the point $\xi$ corresponding to the trial orbit, i.e. 
\begin{equation}
\Delta s^2 = (\xi^1)^2 + \cdots + (\xi^6)^2.
\end{equation}

Had we carried on to the second order transformation using the $B\indices{^\mu_\nu_\rho}$ coefficients in Eq.~\ref{eqn:coordtransform_powers}, the above result would extend to a local region. In other words, throughout a neighborhood surrounding $\theta_0$, the expected loss in significance would be simply $\Delta s^2 = (\Delta \xi^1)^2 + \cdots + (\Delta \xi^6)^2$, where $\Delta \xi^\mu$ are the normal coordinate differences between the trial and true orbits, and this would hold to first order anywhere within the neighborhood, not just at $\theta_0$.

In practice, the local coordinates remain flat to a good approximation out to different distances along the different $\xi^\mu$ directions. The coordinate $\xi^6$, corresponding to the smallest eigenvalue of $g$, can be identified with Milani's ``weak direction'' \citep{1999Icar..137..269M,2005A&A...431..729M}. Taking a unit step in $\xi^6$, although it is only predicted to change the relative SNR by $\Delta s^2 = 1$, is a very large step in the original orbital elements, as can be seen from Eq.~\ref{eqn:xi2theta} ($D^{-1}_{66}$ is large). Physically, what is happening is that $\xi^6$ is a nearly degenerate direction in the original orbital parameters, where moving along it barely changes the sky position in the intersected images at all. The change in $\theta^\mu$ for $\Delta \xi^6 =1$ is often large enough that the linear theory breaks down and the image coordinates $\Delta x^i$ change by much more than a PSF size. In contrast, for the ZTF data, it is usually possible to move by thousands of unit steps along $\xi^1$ and $\xi^2$ and the shifts $\Delta x^i$ remain confined to a compact region of the sky.

\begin{figure*}
\begin{center}
\includegraphics{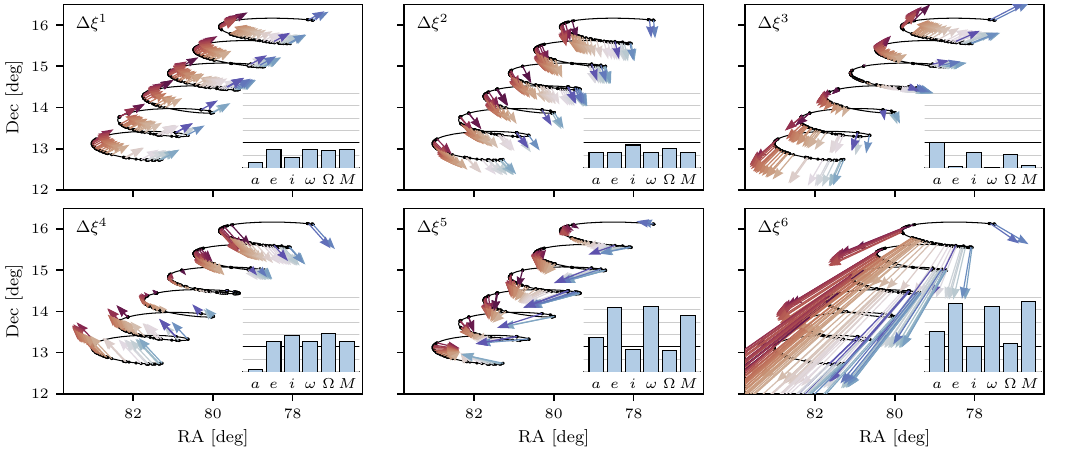}
\end{center}
\caption{\label{fig:DskyDlocalcoords} Illustration of local coordinates for orbital parameter space $\xi^\mu$ near the orbit of numbered minor planet 612911, as seen over six years of ZTF imaging. Each small dot and attached arrow correspond to a time when the object was present in a ZTF image. The arrow shows the shift in the orbital path corresponding to a change $\Delta \xi^\mu = 1$ holding the other local coordinates fixed (for visibility, arrows are magnified by a factor of 1000 and are colored according to time of year). The sets of arrows for the different $\xi^\mu$ form a basis for exploring the local region of orbital parameter space. The inset in each panel shows the combination of Kepler element shifts corresponding to a change $\Delta \xi^\mu=1$ in the local coordinate. The height of a bar is the logarithm of the absolute value of the Keplerian element shift $\Delta \theta^\nu$ in units of the Kepler element's natural length scale $\ell^\nu$ (see Eqs.~\ref{eqn:xi2theta} and~\ref{eqn:elemscales}). The horizontal black line corresponds to $\lvert \Delta\theta^\nu / \ell^\nu\rvert = 1$ and each gray line is a factor of 10.}
\end{figure*}
Figure~\ref{fig:DskyDlocalcoords} illustrates the local coordinates for the numbered minor planet 612911, a high-inclination outer solar system object currently 57~au from the Sun. The helical path shows its sky position over the duration of the ZTF survey. The small dots with attached arrows correspond to times when 612911 was present in a ZTF exposure. The arrows show the shift in sky coordinates corresponding to a change in each of the local coordinates by $\Delta \xi^\mu=1$, holding the others constant. The arrows are magnified by a factor of 1000 for visibility. In the linear approximation, $\Delta \xi^\mu = 1$ leads to $\Delta s^2 = 1$, which corresponds to a shift in the sky of about a PSF size in every image.

Each set of arrows corresponds to a column of the $U$ matrix in the SVD decomposition of the Jacobian (Eq.~\ref{eqn:SVD}). These columns are a basis for describing all possible small changes in the orbit of this object. The various ``eigenorbits'' have nicely distinct behaviors. For example, $\xi^1$ and $\xi^2$ are simple translations of the orbital path, $\xi^4$ rotates the path, and $\xi^3$ and $\xi^5$ are stretches roughly parallel and perpendicular to the long-term sky motion. In contrast, a similar plot of the sky shifts caused by changing each Kepler element separately shows that most of the Kepler elements are highly degenerate with one another. Except for semi-major axis, changing the other Kepler elements appears to simply translate the orbit on the sky (similar to $\xi^1$ and $\xi^2$ above). The local coordinates isolate the subtle ways that these translations are actually not uniform over the duration of the observations. In our experience, the translational and rotational ``modes'' always occur, but the stretching modes for some orbits are not always obviously separated as parallel and perpendicular to the long-term motion.

The inset in each panel of Fig.~\ref{fig:DskyDlocalcoords} shows the linear combination of the different Kepler elements corresponding to the local coordinate (Eq.~\ref{eqn:xi2theta}). In the $\xi^1$ panel, for example, the bar heights show the absolute value of the Kepler element shifts $\Delta\theta^\mu$ corresponding to $\Delta \xi^1 = 1$. The $\Delta\theta^\mu$ are normalized by their natural length scales $\ell^\mu$ (Eq.~\ref{eqn:elemscales}) and the bars are plotted in logarithmic scaling. The horizontal black line corresponds to $\lvert \Delta\theta^\mu / \ell^\mu\rvert = 1$ and the gray lines are steps by factors of 10. The first three local coordinates correspond to Kepler element shifts that would be expected to shift the sky position by about 1 pixel in each image ($\Delta\theta^\mu \approx \ell^\mu$). The next three local coordinates correspond to much larger changes in Kepler elements that nonetheless conspire to keep the on-sky changes small. The last local coordinate $\xi^6$ corresponds to a near-exact degeneracy in orbital elements: $\Delta \xi^6 = 1$ is expected to be a barely distinguishable change in sky coordinates but involves changing the Kepler elements by many thousands of times the scale that would cause a shift by one PSF. This is a weak direction where the linear theory breaks down and the actual change in sky coordinates will cause more than a $\Delta s^2=1$ drop in detection significance, as seen by the longer arrows.

\subsection{Use of the local coordinates to construct a lattice of trial orbits}

The local coordinates can be used to build a lattice of trial orbits for digital tracking. The spacing between points in $\xi$-coordinates should be such that the maximum squared distance from any point to the closest grid point is $\Dsmax$. This ensures that the worst-case relative reduction in SNR due to mismatch between the trial and true orbits is $\exp(-\Dsmax)$.

We find it is possible to grow local coordinate patches around a seed orbit $\theta_0$ by moving along each $\xi^\mu$ coordinate until the metric distance $\Delta s^2$ reaches some value (e.g. $\Delta s^2 = 3000$). However, if a direction $\xi^\mu$ is a weak direction, the extent of the patch is just the $\Delta \xi^\mu$ (which will be less than 1) such that $\Delta s^2=1$. The result is a rectangular region in $\xi$-coordinates (corresponding to a hyperparallelogram in the original orbital elements) that can be either filled with a lattice of trial points, or filled with self-avoiding random points sampled according to the density described in the next section. Details on piecing together local coordinate patches to cover a large region of orbital parameter space will be explored in future work.

\section{The density of trial orbits\label{sec:density}}
The metric leads to a direct answer to the question: how many trial orbits must we test to fully cover a region of orbital parameter space? A region is ``covered'' when every orbit in the region has a squared metric distance at most $\Dsmax$ from the nearest trial orbit. Thus we can be sure that all solar system bodies in the region will be detected with SNR within a fraction $\Dsmax$ of the maximum possible SNR.

In the neighborhood of each orbit $\theta_0$ there is a local coordinate system $\xi^\mu$ in which the metric is Euclidean (Sec.~\ref{sec:normalcoords}). The placement of trial orbits is a packing problem in these local coordinates. There must be roughly one trial orbit per sphere of squared radius $\Dsmax$ in $\xi$-space. An infinitesimal volume $dV_\xi$ in $\xi$-space maps to the infinitesimal volume $dV_\theta$ back in the original orbital elements, where $dV_\theta = |g(\theta_0)|^{-1/2} dV_\xi$ and $|g(\theta_0)|$ is the determinant of the metric tensor in $\theta$-coordinates (this can be derived from Eq.~\ref{eqn:linear_euclidean_transf_matrixform}).

The volume of a $d$-dimensional Euclidean ball with radius $r$ is $V_d(r) = c_d r^d$, where $c_d$ is a constant prefactor (in six dimensions $V_6(r)= \frac{\pi^3}{6} r^6$; the prefactor is roughly 5 for $d=4$ to $7$). Therefore, in $\theta$-space we need one trial orbit per volume $dV_\theta = |g(\theta_0)|^{-1/2} c_d (\Dsmax)^{d/2}$. Equivalently, the required density of trial orbits in orbital parameter space is
\begin{equation}
\rho(\theta) = \frac{|g(\theta)|^{1/2}}{c_d \, (\Dsmax)^{d/2}}. \label{eqn:densityallfactors}
\end{equation}
We define
\begin{equation}
\trialdensity(\theta) = |g(\theta)|^{1/2}, \label{eqn:trialdensitydef}
\end{equation}
to be the benchmark trial orbit density since $\Dsmax$ will usually be chosen to be of order 1. This density is a generally covariant quantity. Calculating the number of trial orbits by integrating $\rho(\theta)$ over a finite volume of phase space will give the same result for any equivalent parameterization of orbits (e.g. Keplerian elements,  equinoctal elements, Cartesian position and velocity vectors at an epoch, etc).

\subsection{Validation of trial orbit density in a blind search\label{sec:blindsearchexperiment}}

We tested the density formalism by performing blind digital tracking searches for two known minor planets in six years of ZTF data (see Sec.~\ref{sec:dataprep} for a description of the data). One of the authors selected distant objects from the Minor Planet Center, queried Horizons for their ``true'' orbital elements (both had uncertainty code 1), and then defined search regions in phase space enclosing the true orbits. Search regions were 6-d rectangular boxes in Keplerian elements where the width in each element $\theta^\mu$ was chosen to be $\sim 600$ times the element's natural length scale $\ell^\mu$ (Eq.~\ref{eqn:elemscales}). This box size is such that the orbits in the search region collectively intersect several thousand ZTF images, which fit into memory on a single compute node. Each search box was then shifted so that the true orbit ended up at a uniformly random location within the box. The density $\rho$ (Eq.~\ref{eqn:densityallfactors}) was calculated and used to predict the number of trial orbits required to discover the hidden object. Specifically, density was calculated at the $3^6$ combinations of lower bound, upper bound, and midpoint along each dimension and then multiplied by the volume of the box. The standard deviation among the density samples was around 15\% and we used the mean to estimate the required number of orbits. The box boundaries and number of trial orbits were then passed to other members of the team who performed the search with our high-performance computing (HPC) digital tracking software \citep{2025A&C....5300987G}. The code draws trial orbits using a space-filling Sobol sampling method, which ensures fairly uniform coverage of the search region.

Both of our experiments were successful in detecting the hidden minor planets. We detail the results for the second one, (341520) Mors-Somnus. This object intersects 780 ZTF images with a predicted median magnitude of $V=21.5$, just fainter than the limiting magnitude for a single image (the Minor Planet Center has no reported observations of Mars-Somnus from ZTF or Palomar). It was predicted that $2.3 \times 10^9$ trial orbits would be needed to detect it at $\Delta s^2 < 1$ and twice this number were trialed by the HPC code, meaning we expect about 2 trial orbits to have SNR within a factor $\exp(-1)$ of $\SNRmax$ corresponding to the true orbit. After the search, the SNR corresponding to the true orbit was measured to be $\SNRmax = 60.3$, and there were 4 trial orbits with $\SNR > 22.2$ with the highest being 46.1.

\begin{figure}
\begin{center}
\includegraphics{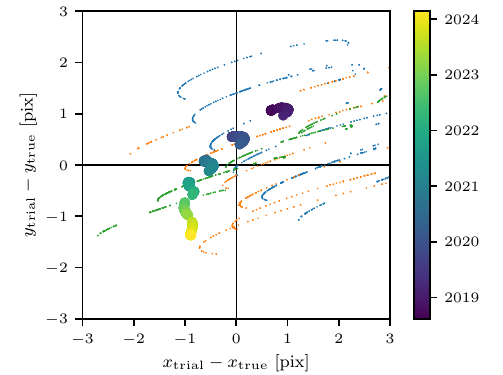}
\end{center}
\caption{\label{fig:blindsearchpixcoords} Recovery of (341520) Mors-Somnus in a blind search experiment. The large circles show the offsets in image coordinates between the trial orbit with the highest observed SNR and Mors-Somnus's true orbit. Each circle corresponds to one of the 780 intersections with ZTF (color shows the date). This trial orbit gave a detected at $\SNR=46$, whereas stacking along the true orbit gives $\SNRmax=60$. The small sets of dots show the next three top trial orbits having $\SNR=31$ (blue), 28 (orange), and 24 (green).}
\end{figure}
Figure~\ref{fig:blindsearchpixcoords} confirms that Mors-Somnus was indeed recovered in the blind search. The large circles show the offset in pixel coordinates between the orbit with the highest SNR and the true orbit. The top trial orbit stays within about 1 pixel over the course of the six-year observations. The other sets of points in this figure correspond to the other top 3 trial orbits. Numerically optimizing the SNR starting at the best trial orbit converges to the true orbit. There are many trial orbits which come close to the true orbit in a subset of intersections. This is reflected in the falloff of significance with $\Delta s^2$ (see Fig.~\ref{fig:sedna_SNR_vs_a} and the discussion after Eq.~\ref{eqn:SNRmetricexpform}). In this experiment there were 345 trial orbits with $\SNR > 10$ and 1800 with $\SNR > 7$. We did not explore whether optimizing SNR from these starting points converges to the true orbit, but a two-step procedure of trial-and-error search at a lower density, followed by optimization of high-SNR orbits may be efficient.

\section{Application to linear motion\label{sec:linearmotion}}
In this section we apply the metric, local coordinates, and trial density to the simple case of objects moving with constant velocity in two dimensions. This is the motion model used for most digital tracking and can be used for stacking observations over short time baselines.

There are four orbital elements in this case: the position $(x_0,y_0)$ at the epoch $t_0$ and the (constant) velocities $(v_x,v_y)$ in the two directions, i.e. \mbox{$\theta^\mu = (x_0,v_x,y_0,v_y)$}. For motion in the sky tangent plane, the velocities are the components of proper motion (e.g. in arcsec/day). Observations take place at the times $t_1, \ldots, t_N$ measured from the epoch. At these times the object is located at
\begin{equation}
\begin{pmatrix}
x_1 \\ y_1 \\ \vdots \\ x_N \\ y_N
\end{pmatrix} =
\begin{pmatrix}
x_0 + v_x t_1 \\ 
 y_0 + v_y t_1 \\
 \vdots \\
  x_0 + v_x t_N \\ 
 y_0 + v_y t_N
\end{pmatrix}. \nonumber
\end{equation}

Therefore, the Jacobian of the transformation from orbital elements to sky positions is
\begin{equation}
J^i_\mu = 
\begin{pmatrix}
\frac{\partial x_1}{\partial x_0} & \frac{\partial x_1}{\partial v_x} & \frac{\partial x_1}{\partial y_0} & \frac{\partial x_1}{\partial v_y} \\
\vdots&\vdots&\vdots&\vdots\\
\frac{\partial y_N}{\partial x_0} & \frac{\partial y_N}{\partial v_x} & \frac{\partial y_N}{\partial y_0} & \frac{\partial y_N}{\partial v_y}
\end{pmatrix}
=
\begin{pmatrix}
1 & t_1& 0  & 0 \\
0 & 0   & 1  & t_1\\
\vdots&\vdots&\vdots&\vdots\\
1 & t_N & 0 & 0 \\
0 & 0    & 1 & t_N
\end{pmatrix}. \nonumber
\end{equation}

For simplicity, assume each image has the same pixel noise level $\sigma^2$, a Gaussian PSF with standard deviation $b$, small pixels, and that objects maintain constant brightness over time. Then the $c_i$ weights in the image-space metric $\eta_{ij}$ (Eq.~\ref{eqn:eta_def}) are all $1/N$ and $\eta_{ij}$ is simply a constant diagonal matrix with entries $1/(4Nb^2)$.

The metric in orbital parameter space (Eq.~\ref{eqn:metricdef}) is
\begin{equation}
g_{\mu\nu} = J^i_\mu \eta_{ij} J^j_\nu 
= \frac{1}{4b^2}
\begin{pmatrix}
1                        & \langle t \rangle   & 0                          & 0 \\
\langle t \rangle & \langle t^2 \rangle& 0                           &0 \\
0                        & 0                          & 1                           & \langle t \rangle \\
0                        & 0                          & \langle t \rangle     &  \langle t^2 \rangle 
\end{pmatrix}, \nonumber
\end{equation}
where $\langle t \rangle = \frac{1}{N}\sum_i t_i$ and $\langle t^2 \rangle = \frac{1}{N}\sum_i t_i^2$. If we choose the epoch $t_0$ to be the mean observation time, then $\langle t\rangle=0$ and $\langle t^2 \rangle$ is the variance of the distribution of observation times, a measure of the survey baseline. The metric is then simply
\begin{equation}
g_{\mu\nu} = \frac{1}{4b^2}\Diag\left(1, \langle t^2 \rangle, 1, \langle t^2 \rangle\right). \nonumber
\end{equation}

The metric does not depend on location in orbital elements space. Phase space is flat and there is a single set of normal coordinates $(\xi^1, \ldots \xi^4)$ for which the metric is everywhere Euclidean. The transformation is simply
\begin{align*}
x_0 &= 2b \xi^1 & v_x &= \frac{2b}{\sqrt{\langle t^2 \rangle}} \xi^2 \\
y_0 &= 2b \xi^3 & v_y &= \frac{2b}{\sqrt{\langle t^2 \rangle}}  \xi^4. 
\end{align*}

In a digital tracking search, the trial orbit lattice could be a square grid in the 4-d $\xi$-space. The grid spacing should be such that the maximum squared distance from any point to the closest grid point is $\Dsmax$. This ensures that the worst-case relative reduction in SNR due to mismatch between the trial and true orbits is $\exp\left(-\Dsmax\right)$ (Eq.~\ref{eqn:SNRmetricexpform}). If the $\xi$-grid has spacing $\delta$ in all four dimensions, the worst-case scenario is when the true orbit is at the center of a grid cell, giving $\Dsmax = 4 (\delta/2)^2 = \delta^2$. For $\Dsmax = 1$ (i.e. the tolerable reduction of significance is a few 10s of percent) we should set the grid spacing to be \mbox{$\Delta \xi^\mu = \delta = 1$} in every dimension.

Translating back to the original orbital elements, the spacing between trial initial positions would then be $\Delta x_0 =\Delta y_0 = 2b$ and the spacing between trial velocities would be $\Delta v_x = \Delta v_y = 2b / \sqrt{\langle t^2 \rangle}$. This is similar to the spacing heuristically derived by \citet[][Eq.~4]{2015AJ....150..125H}, who set the velocity spacing to be $\sqrt{2}$ times the maximum acceptable blur divided by the time spanned by the observations. \citet{2009AJ....137.4400L} also note that the relevant timescale for measuring proper motion is the variance of the observation times rather than the total time span.

The benchmark density of trial orbits (Eq.~\ref{eqn:trialdensitydef}) needed for digital tracking to discover linear motion is simply
\begin{equation}
\trialdensity = \frac{\langle t^2 \rangle}{64 b^4}. \label{eqn:linearmotiondensity}
\end{equation}
The required number of trial motions is just the product of density and the volume of phase space to be searched. If the sky area surveyed is $\skyarea$ and the goal is to find sources moving in any direction with proper motion up to $\vmax$, the phase space volume is $\skyarea \pi \vmax^2$. Combining these factors and using Eq.~\ref{eqn:densityallfactors} with $c_4=\pi^2/4$, the total number of trials needed is
\begin{equation*}
N_\mathrm{trial} =  \frac{\skyarea\vmax^2 \langle t^2 \rangle}{16 \pi b^4(\Dsmax)^2}.
\end{equation*}

For the simple assumption of constant survey cadence, the time variance is $\langle t^2 \rangle = T^2 / 12$, where $T$ is the total survey time. Putting in units, the number of trials needed per sky area surveyed is \citep{mcgill2025inprep}
\begin{equation}
\frac{N_\mathrm{trial}}{\skyarea} =  \frac{6.6 \times 10^{5}}{\deg^2} \frac{\left(\dfrac{\vmax}{1~\mathrm{arcsec/day}}\right)^2 \left(\dfrac{T}{\mathrm{day}}\right)^2}{\left(\dfrac{\mathrm{PSF~FWHM}}{\mathrm{arcsec}}\right)^4 \left(\Dsmax\right)^2}. \label{eqn:Ntrialslinear}
\end{equation}

\section{Application to the Zwicky Transient Facility\label{sec:ZTF}}
The results in this paper are in preparation for a digital tracking search using the Zwicky Transient Facility survey \citep{2019PASP..131a8002B,2019PASP..131f8003B}. In this section we calculate the sensitivity limit when stacking images across the multi-year survey and the density of trial orbits needed to carry out such a search.

\subsection{Data preparation and orbit sampling\label{sec:dataprep}}
We obtained all $r$ and $g$-band difference imaging produced by ZTF \citep{2019PASP..131a8003M,https://doi.org/10.26131/irsa539} over approximately six years from March 20, 2018 to February 29, 2024. This comprises 849,750 30-second exposures, stored in about 50~million ccd-quadrant FITS files, each containing a $3072\times3080$-pixel image covering a sky area of $\approx 0.75~\mathrm{deg}^2$. The survey covers the entire sky above declination $-30^\circ$. The limiting magnitude for a $5\sigma$ point source detection in individual $r$-band images is $20$ to $21.3$ (16th to 84th percentile) and similar in $g$ band. The median seeing is 2.0 and 2.2~arcsec in $r$ and $g$ images (FWHM) using a fiducial pixel scale of 1.012~arcsec/pix. ZTF observations are ongoing but we will refer to the ``ZTF survey'' to mean the 6-year span we have on disk.

To assess sensitivity to a diversity of minor planet populations we generated $2^{27}$ (about 130 million) orbits over the 6-d space of Keplerian orbital elements using a space-filling Sobol sampler\footnote{We used the \texttt{scipy} (v1.15.1) implementation \texttt{scipy.stats.qmc.Sobol} with default settings.}. Sampling was log-uniform in semi-major axis between 5 and 1000~au, uniform in eccentricity between 0 and 1, uniform in $\cos\left(\text{inclincation}\right)$ between $0^\circ$ and $180^\circ$, and uniform between $0^\circ$ and $360^\circ$ in argument of perihelion, longitude of ascending node, and mean anomaly. We do not explore semi-major axes less than 5~au because, as shown below, the density of required trial orbits increases dramatically in the inner solar system.

For each sample orbit $\theta$ we identified the ZTF intersections (the images that would contain an object on that orbit) and computed the pixel coordinates within those images using the WCS parameters for the TAN projection \citep{2002A&A...395.1077C} in the FITS header (ignoring the TPV distortion coefficients). Orbits are modeled as Kepler ellipses around an inertial Sun, neglecting perturbations from the planets\footnote{This simplification does not affect the results of this paper on limiting magnitude and the density of distinguishable orbits. For the outer solar system, Keplerian motion is also sufficient for the actual stacking search, maintaining arcsec precision over several years. However, a more complete force model must be used to perform a stacking search in the inner solar system.}. We obtained the Jacobian $J^i_\mu$ of the transformation from $\theta$ to pixel coordinates (Eq.~\ref{eqn:jacobiandef}) by numerical differentiation\footnote{We used the finite difference package \texttt{numdifftools} \citep{numdifftools,AdaptiveRobustNumericalDifferentiation}, the \texttt{numdifftools.Jacobian} function with \texttt{order=4}, \texttt{num\_steps=5}, and an adaptive procedure for setting $\texttt{base\_step}$. The estimated maximum relative error in any element of the Jacobian was $\sim 10^{-7}$ for the median orbit, and larger than 1\% in about 1 in $10^5$ sample orbits.}.

The density of trial orbits and the expected detection significance require the (relative) object flux in each image, the image PSFs, and pixel noise variances. The PSF is assumed Gaussian (Eq.~\ref{eqn:2dgaussianpdf}) with $b_i$ calculated from the \texttt{SEEING} parameter in the FITS header. A robust pixel noise variance $\sigma_i^2$ is calculated with an iteratively sigma-clipped MAD (median absolute deviation from the median) estimator applied to the image's pixel values. The predicted flux $\tA_i$ is the light curve computed in the H,G magnitude system for V band, color-corrected to $r$ or $g$ using colors we derived for Makemake\footnote{Makemake's normalized reflectance spectrum is taken from \citet[][Fig.~1]{2020MNRAS.497.5473A} and multiplied by the solar spectrum at zero airmass \citep{ASTM_E490-00AR19} to obtain an unnormalized spectrum. Its $AB$ magnitudes are then found for the Pan-STARRS1 bandpasses \citep{2012ApJ...750...99T} used by ZTF. The resulting colors are $V-r=0.24$ and  $V-g=-0.37$.}, and then converted to image units using the zero point magnitude in the FITS header\footnote{\texttt{MAGZP} in the header is adjusted using the \texttt{CLRCOEFF} coefficient and Makemake's $g-r$ color.}. We use Makemake's color to optimize for the outer solar system but do not expect the color correction to have a strong effect on the sensitivity of the search.

\begin{figure}
\begin{center}
\includegraphics{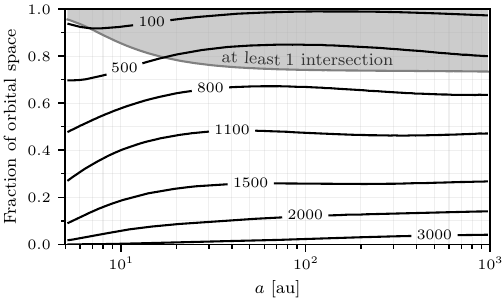}
\end{center}
\caption{\label{fig:numintersections_vs_a} The fraction of orbital parameter space intersecting the ZTF survey in various numbers of images. The boundary of the gray shaded region shows the fraction of orbits that will be captured in at least one ZTF image (orbits are sampled as described in Sec.~\ref{sec:dataprep} and binned in semi-major axis). The black curves show the distribution of the number of ZTF intersections for the orbits that have at least one intersection. Specifically, of all orbits that are present in ZTF, the curves shows the fraction that have greater than 100, 500, 800, etc intersections. For example, the median outer solar system orbit observed by ZTF will be present in a stack of about 1100 images.}
\end{figure}
Figure~\ref{fig:numintersections_vs_a} shows the distribution of the number of ZTF images intersected by the population of orbits as a function of semi-major axis. The boundary of the gray shaded region gives the fraction of sampled orbits that have at least one intersection with ZTF. For high semi-major axes, a solar system object's sky position is roughly constant over 6~years and the fraction of orbits with a ZTF intersection just reflects the three quarters of the sky covered by the survey. At low semi-major axes, nearly all bodies are moving fast enough that they will enter the northern hemisphere at some point over the 6-year baseline and be captured by ZTF. The black curves in the figure are conditional on orbits that have at least one intersection. They show the fraction of such orbits with greater than 100, 500, etc intersections in ZTF. Nearly all moving objects observable with ZTF will typically be seen in hundreds to a few thousand images. Image stacking sensitivity, scaling as $\SNR \propto \sqrt{N}$ for $N$ intersections, means there is a substantial benefit from stacking in years-long, wide-area surveys like ZTF --- assuming the computational demands can be met.

\subsection{Sensitivity limit of a stacking search\label{sec:ztfdepth}}
For each sample orbit, the absolute magnitude corresponding to an $\SNRmax=10$ detection, denoted $\Hten$, is computed with Eq.~\ref{eqn:SNRmaxexact}. This represents the faintest object (on the given orbit) a stacking search can be expected to discover. For this limiting magnitude we also compute the expected SNR and apparent V magnitude in the individual exposures making up the stack. We choose $\SNRmax=10$ instead of, say, 5 because the naive significance (sometimes called local significance) must be reduced due to the huge number of orbits that are tested. However, it is easy to scale our sensitivities to other thresholds: expected significance is proportional to an overall scaling in the source flux so the absolute magnitude corresponding to, say, $\SNRmax=60$ would be $H_{60} = \Hten -2.5 \log_{10}(60/10)$.

\begin{figure}
\begin{center}
\includegraphics{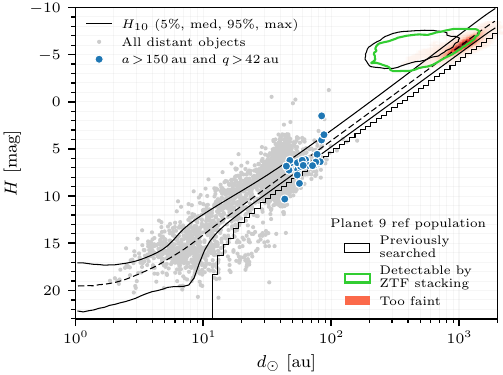}
\end{center}
\caption{\label{fig:Hdepth} Depth of a ZTF stacking search as a function of a solar system body's heliocentric distance at the midpoint of the survey. The thin black lines are upper limits on an object's absolute magnitude ($\Hten$) that yield an $\SNR \geq 10$ stacked detection. From top to bottom they show the 5th percentile, median (dashed), 95th percentile, and maximum (jagged) $\Hten$ among all orbits that have at least one intersection with ZTF. In general, objects with $H<\Hten$ will be detectable by stacking. The gray points show the currently known population of distant objects, and the blue markers highlight the population used to infer the orbit of the hypothesized Planet~9. The three regions in the upper right show the possibilities for Planet~9 itself, based on the reference models of \citet{2022AJ....163..102B}. The thin black contour surrounds 99\% of the sample Planet~9's that would have already been detected by ZTF, Pan-STARRS, or DES (78\% of the reference population). The green 99\%-contour shows candidates that have been previously undetectable but could be seen at $\SNR>10$ with ZTF stacking (18\% of the population). The red heat map shows the remaining 1\% of orbits that have at least 10 ZTF intersections but are too faint to be detected by stacking (an additional $2.7\%$ of the reference population is undetectable because it falls outside the ZTF footprint).}
\end{figure}

An object's distance is the dominant factor in $\Hten$. Figure~\ref{fig:Hdepth} shows how $\Hten$ scales with heliocentric distance at the midpoint of the ZTF survey, which we denote $\dsun$. For each bin in $\dsun$ the thin black lines show the 5th percentile, median (dashed line), and 95th percentile of $\Hten$ among all orbits in that bin which have at least one ZTF intersection. The jagged line corresponds to the orbit with the largest value of $\Hten$, i.e. the most favorably observed orbit.

The variation in depth among orbits at equal distance is mainly due to differing numbers of intersections. For a moving object with absolute magnitude $H$ that appears in $N$ images, the detection significance scales approximately as (see Eq.~\ref{eqn:SNRmaxsmallpix})
\begin{equation}
\SNR \propto \frac{10^{-H/2.5}}{\dsun^4} \sqrt{N}, \nonumber
\end{equation}
which implies
\begin{equation}
\Hten = -10 \log_{10} \dsun +1.25 \log_{10} N + \mathrm{const}. \nonumber
\end{equation}
We fit a linear relation between $\Hten$ and either $\log_{10}\dsun$ alone or $\log_{10}\dsun$ and $\log_{10}N$ to the orbits with $\dsun > 20~\au$ (half the sample is used for fitting and the other half to compute the root mean square of the residuals). The results are
\begin{align*}
\Hten &= -9.91 \log_{10}\frac{\dsun}{\au} + 23.83 & \mathrm{rms}= 0.60& \\
\Hten &= -9.91 \log_{10}\frac{\dsun}{\au} + 1.64 \log_{10} N + 18.97 & \mathrm{rms}=  0.17&.
\end{align*}
Including the number of intersections soaks up most of the variation. The remainder is due to variation in seeing, noise level, and zero point among the individual images. Interestingly, the best fit coefficient for $\log_{10}N$ is 1.64 rather than the expected 1.25, meaning that depth seems to scale somewhat more favorably than $\sqrt{N}$, perhaps because $N$ is correlated with sky position and therefore image quality in some way.

For context and motivation, Fig.~\ref{fig:Hdepth} also highlights several solar system populations. The gray points show the Minor Planet Center's distant object list\footnote{6,441 objects as of July 18, 2025.} with the distant Kuiper Belt objects shown as blue circles. 

The distant KBOs are selected according to the same criteria as in \citet{2021AJ....162..219B} ($a > 150~\au$ and $\mathrm{perihelion}>42~\au$). Apparent clustering in this population has been used to hypothesize the existence of a large planet (Planet~9) at many hundreds of $\au$ \citep{2016AJ....151...22B,2019PhR...805....1B} (but see also \citet{2017AJ....154...50S} for a cautionary view). These outer KBOs pile up near the faint end of what ZTF stacking can detect. The benefit of stacking will be in finding these objects over an enormously larger sky area than has been explored at this depth (see Fig.~\ref{fig:Vmagdepth}). Enlarging this population is important for the Planet~9 hypothesis and would alleviate potential selection bias of previous narrow-field observations \citep[see, e.g.][]{2017AJ....154...50S}.

Finally, the top right of Fig.~\ref{fig:Hdepth} shows possibilities for Planet~9 itself. This reference population is described by \citet{2021AJ....162..219B}, who sampled a posterior distribution of Planet~9 orbits and masses given the orbits of the known distant KBOs (\cite{2025ApJ...978..139S} obtained a somewhat different distribution). The reference population also indicates whether each sample Planet~9 would have been detected by searches for linked, single-epoch detections performed in ZTF, the Dark Energy Survey (DES), and Pan-STARRS1 \citep{2022AJ....163..102B,2022AJ....163..216B,2024AJ....167..146B}. The $H$ mag of each Planet~9 candidate is computed from the V mag listed in the table and its heliocentric distance (assuming zero phase angle and equal geocentric and heliocentric distances). Then, for each Planet~9 orbit, the limiting $\Hten$ for ZTF stacking is found as described above. The black contour encloses 99\% of the Planet~9 samples that are already ruled out by ZTF, DES, and Pan-STARRS1. This constitutes 78\% of the reference population. The green contour encloses 99\% of the remaining Planet~9 samples that would be detectable at $\SNR=10$ in a stacked ZTF search (18\% of the reference population). Stacking pushes to both fainter fluxes and to further distances compared to the parameter space already explored. The red heat map shows the 1\% of Planet~9 reference orbits that intersect at least 10 ZTF images but at magnitudes too faint to be detected by stacking. A remaining 2.7\% of allowed orbits intersect fewer than 10 images and are not shown. Stacking in ZTF, therefore, has the ability to detect or rule out the large majority (80\%) of the remaining parameter space for Planet~9.

\subsection{Density of trial orbits in ZTF\label{sec:ztfdensity}}
\begin{figure}
\begin{center} 
\includegraphics{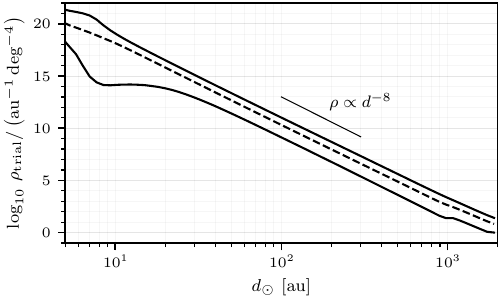}
\end{center}
\caption{\label{fig:density} The density at which Keplerian orbital elements must be tested to discover solar system bodies in a 6-year ZTF stacking search. The curves show the 5th, median, and 95th percentiles over orbital parameter space conditioned on $\dsun$, an object's heliocentric distance at the midpoint of ZTF.}
\end{figure}
Equations~\ref{eqn:eta_def}, ~\ref{eqn:metricdef}, and ~\ref{eqn:trialdensitydef} are used to compute the density of trial orbits $\trialdensity(\theta)$. Figure~\ref{fig:density} shows the distribution of trial orbit density as a function of heliocentric distance of the orbit $\dsun$. As with $\Hten$, density is determined primarily by $\dsun$ and the curves show the 5th percentile, median, and 95th percentile of our sampled orbits within each $\dsun$ bin. The units of density are $1/(\text{volume of Keplerian orbital elements space})$, i.e. $\mathrm{au}^{-1}\deg^{-4}$. 

\begin{figure}
\begin{center}
\includegraphics{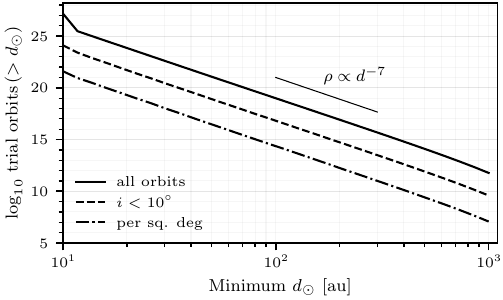}
\end{center}
\caption{\label{fig:integrateddensity} The number of trial orbits required to fully search the 6-year ZTF data. The curves show the integral of the required orbit density $\rho(\theta)$ over all orbits for which the object is beyond heliocentric distance $\dsun$ at the midpoint of ZTF. The solid curve is for the full range of bound orbits, dashed is restricted to inclinations less than $10^\circ$, while dot-dashed gives the number of orbits needed to saturate each square degree patch of sky.}
\end{figure}
The integral of $\trialdensity$ over a finite orbital region is the total number of distinguishable orbits that must be tested to fully search that region. While the trial orbit density depends on the parameterization of orbital phase space, its integral is independent of phase space coordinates and directly quantifies the computational scale of the digital tracking problem.

We compute $\int \trialdensity(\theta) \,d^6 \theta$ by quasi-Monte Carlo integration using our $2^{27}$ sample orbits (including a factor $a/\sin(i)$ in the integrand to adjust for the non-uniform orbit sampling in semi-major axis and inclination). Figure~\ref{fig:integrateddensity} shows the total number of trial orbits needed to fully search various regions of orbital phase space using the 6-year ZTF data (using Eq.~\ref{eqn:densityallfactors} with $\Dsmax=1$, $c_6=\pi^3/6$). For each value of $\dsun$, the curves give the number of trial orbits needed to find objects at heliocentric distance $\dsun$ and beyond. The solid line includes all bound orbits (i.e. the full range for each Keplerian element), while the dashed line shows orbits with inclinations less than $10^\circ$. The dot-dashed line gives the number of trial orbits needed to ``fill'' a square degree patch of sky. It is computed by restricting the integration region to orbits for which the object is in a particular patch of sky at the midpoint of ZTF. The number of required orbits grows linearly with the area of the patch and there is minimal dependence on the location of the patch within the ZTF footprint. We checked the accuracy of the integration by dividing our sample orbits into 32 subsamples and re-calculating the results, finding negligible variation except at the lowest value of $\dsun$.

The number of required trial orbits scales extremely rapidly with inverse distance, approximately as $\trialdensity \propto \dsun^{-7}$. This harsh scaling makes a fully coherent 6-year blind stacking search for inner solar system objects prohibitively expensive. Even the outer solar system will be a challenge for the world's largest computers, at least with the brute-force, trial-and-error approach. We discuss possible methods to ease the computational burden in Sec.~\ref{sec:tradeoffs}.

\section{Discussion\label{sec:discussion}}
Multi-year digital tracking can increase the sensitivity of a moderately sized survey telescope to that of a world-class instrument. With ZTF, for example, stacking ${N \approx 10^3}$ images essentially improves the detection threshold by 3 to 4 magnitudes over the nominal $r=21$ single-exposure limit. This stacked depth is on par with individual exposures of the Rubin Observatory's Legacy Survey of Space and Time \citep[LSST;][]{2019ApJ...873..111I}, which will detect point sources at $r \approx 24.6$.

\begin{figure}
\begin{center}
\includegraphics{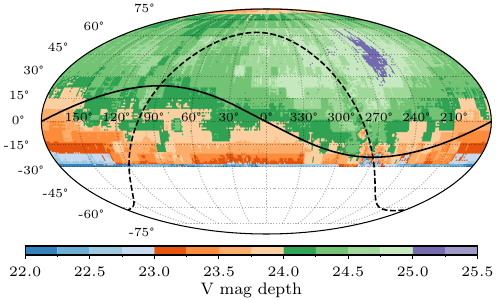}
\end{center}
\caption{\label{fig:Vmagdepth} Median V magnitude in individual ZTF exposures leading to an $\SNR=10$ detection in a stacked search. For comparison, the median $5\sigma$ magnitude limit for source detection in single ZTF images is around 21 (similar in $r$ and $g$), while for Rubin LSST the single-image depth will be $r\approx 24.6$. The solid and dashed black lines are the ecliptic and galactic planes. This figure includes all sampled orbits that would appear in at least one ZTF image. Declinations below $-30^\circ$ are masked, which erases the very small fraction of orbits that enter the ZTF footprint after, or exit before, the midpoint of the survey.}
\end{figure}

Figure~\ref{fig:Vmagdepth} shows the effective depth ZTF achieves over the wide swath of sky it surveys. Solar system bodies with apparent V magnitude brighter than the limit shown will be detected at $\SNR=10$ in a stacked search. The depth varies somewhat across the sky because of ZTF's observing pattern. To produce the figure, for each of our sample orbits, we scale its absolute magnitude so that the stacked SNR is 10 and then find the median V magnitude such an object would have in the ZTF images it intersects. Orbits are then binned spatially into $1^\circ \times 1^\circ$ regions according to their sky location at the midpoint of the survey. For each spatial bin, the color shows the median over the orbits in that bin. We use our full orbit sample, but the figure is unchanged if only orbits with semi-major axes greater than 10, 20, or $100~\au$ are included.

The implication is that surveys like ZTF or Pan-STARRS \citep{2002SPIE.4836..154K} can provide a northern hemisphere capability that compliments LSST. This might be used, for example, in pre-discovery searches of LSST alerts \citep{sage2025inprep}. While the required density of trial orbits for near-Earth objects is enormous, the volume of search space would be much reduced if an approximate initial orbit is determined by LSST. It should also be possible to do digital tracking within LSST itself, which would reach unprecedented sensitivity to faint outer solar system objects. The deep drilling fields, in particular, will reach a stacking depth of $r \gtrsim 28.5$\footnote{LSST Baseline Strategy, \url{https://survey-strategy.lsst.io/baseline/index.html}, accessed 2025-09-10.} \citep{2022ApJS..258....1B}. Section~\ref{sec:LSSTlinear} considers digital tracking with Rubin over short time spans.

\subsection{Tradeoff between processing effort and depth\label{sec:tradeoffs}}
\begin{figure}
\begin{center}
\includegraphics{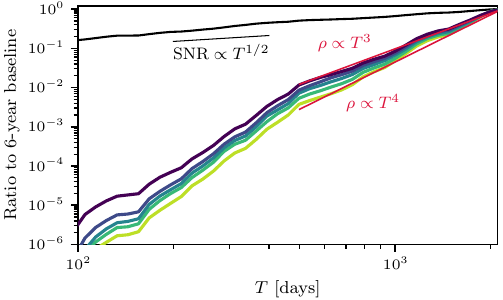}
\end{center}
\caption{\label{fig:snr_density_vs_time} Scaling of depth and trial orbit density of digital tracking with ZTF as a function of the time span $T$ over which images are stacked. The black line shows the detection significance, scaling roughly as $\SNR \propto \sqrt{T}$ as expected for stacking. The colored lines are the median trial orbit density $\trialdensity$ for sample orbits in 5 bins of heliocentric distance $\dsun$ (log-spaced from 10 to 1000~au; distance increases up the plot from green to purple). The density scaling depends on the orbit (thin red lines): distant orbits scale as $\rho \propto T^3$ while those closer in scale as $\rho \propto T^4$. The difference in scaling between SNR and density means that large computational savings can be had at relatively minimal loss of sensitivity.}
\end{figure}
The number of trial orbits that must be tested in wide-area, multi-year digital tracking is clearly enormous (Sec.~\ref{sec:ztfdensity} and Fig.~\ref{fig:integrateddensity}): a six-year northern hemisphere search for all solar system bodies further than 100~au would require $10^{19}$ trial orbits.

One move toward feasibility would be to divide the survey into shorter time segments and perform stacking separately on the images from each segment. To quantify the cost-benefit, we take 2$^{15}$ of our sample orbits (Sec.~\ref{sec:dataprep}) and recalculate the density and expected detection significance as the time span of ZTF is reduced. Specifically, for each time span $T$ we ignore images taken more than $T/2$ from the midpoint of ZTF. Figure~\ref{fig:snr_density_vs_time} shows the results. The black curve is the scaling of expected detection significance $\SNRmax$ with $T$ (the median over all sample orbits at each value of $T$ is shown; there is little variation with orbit). The scaling is roughly $\sqrt{T}$, as expected for image stacking. The colored curves show the scaling of the required trial orbit density $\trialdensity$. The sample orbits are binned by heliocentric distance at the midpoint of ZTF ($\dsun$ used above) into 5 log-spaced bins between 10 and 1000~au and the density ratio compared to $T=6~\mathrm{yrs}$ is found for each orbit and each $T$. The curves show, at each value of $T$, the median ratio for the orbits in a bin. There is some variability in the scaling relation, with orbits with larger $\dsun$ having a shallower density slope ($T^3$ for $\dsun \sim 600~\au$ vs. $T^4$ for $\dsun \sim 15~\au$). The reason for the apparent break at 500~days is also unclear, but perhaps the variations in slope are caused by a change in the effective dimensionality of the motion model.

In any case, the very different scaling between SNR and density means that as the time span of the sub-stacks decreases, the search gets easier ``faster'' than it loses sensitivity. This suggests a hierarchical algorithm for years-long digital tracking. The first step would be to stack over short time spans but with a relatively low detection threshold. Trial orbits or regions of phase space which show promise can then be searched more densely in stacks over longer time spans. This is somewhat reminiscent of the semi-coherent algorithm \citet{2016arXiv161006831S} proposed to search for radio pulsars in long time series.

There are other ways to trade sensitivity for computational cost. Equation~\ref{eqn:linearmotiondensity} shows that for a linear motion model the number of trial orbits increases as the fourth power of the inverse PSF size. For nonlinear digital tracking the scaling is likely to be even steeper: from Eqs.~\ref{eqn:eta_def} and \ref{eqn:metricdef}, for a general 6-parameter motion model the metric $g$ is $b^{-2}$ times a $6\times6$ matrix, so that $\trialdensity = |\det{g}|^{1/2} \propto b^{-6}$. This suggests that increasing the effective PSF size will drastically reduce the number of trial orbits. This can be done, for example, by combining pixels to produce lower-resolution images or convolving images with blurring kernels. Artificially increasing the PSF will decrease significance, but at a much slower rate than $b^{-6}$ ($\SNRmax \propto b^{-1}$ according to Eq.~\ref{eqn:SNRmaxexact}, though that equation does not account for the pixel noise becoming correlated).

\subsection{Digital tracking with Rubin\label{sec:LSSTlinear}}
With a single night of dedicated observation, the Rubin Observatory can conduct an exceptionally deep search for outer solar system objects. Over the course of $T=8~\mathrm{hrs}$, Rubin can take $N\approx 800$ 30-second exposures of a $9.6~\deg^2$ field of view\footnote{Declinations below $-8.3^\circ$ are amenable to 8-hour continuous viewing from Cerro Pach\'on at airmass less than 2, which includes the ecliptic between $201^\circ$ and $339^\circ$ ecliptic longitude. See Appendix~\ref{sec:continuousviewingfield} for details.}. The single-exposure $5\sigma$ magnitude limit in $r$ band is around 24.6, with an effective PSF FWHM of 0.7~arcsec. This means that a stacking search can make an $\SNRmax=10$ detection of a moving object at $r=27.5$.

The feasibility of stacking these observations can be found via Eq.~\ref{eqn:Ntrialslinear}, which is applicable to the outer solar system at this timescale. The number of trial orbits required to test all proper motions less than $21.6~\mathrm{arcsec/day}$ (the mean motion of Neptune) is $N_\mathrm{trial} = 1.4 \times 10^{9}$ for $\Dsmax=1$. Our current searches for nonlinear motion in ZTF are already at this scale and linear stacking is considerably more efficient \citep[e.g.][]{2019AJ....157..119W, 2021AJ....161..282L}.

For $\Dsmax=1$, objects will likely be detected at $\exp(-1)$ of their maximum SNR (Eq.~\ref{eqn:SNRmetricexpform}). Equivalently, objects found at $\SNRobs=10$ will need to be 1.1~mag brighter than the magnitude corresponding to $\SNRmax=10$. The single-night search would thus be complete to $r=26.4$. A depth arbitrarily close to $27.5$ can be achieved by increasing the number of trials. This is about 2 magnitudes shy of the unprecedented depth achieved by \citet{2004AJ....128.1364B} using HST, but covers a sky area around 500 times the size.
\section{Conclusions}
In this paper we assessed the possibilities and challenges of carrying out multi-year, nonlinear digital tracking searches for moving objects with wide-field astronomical surveys. Here we summarize our findings:
\begin{itemize}
    \item The sensitivity and computational difficulty of digital tracking for an arbitrary observing strategy can be quantified with a differential geometry formalism on orbital parameter space. The framework leads to an optimal sampling of trial orbits.
    \item A fully coherent digital tracking search on multi-year, all-sky data has the capability to turn current telescope surveys into the equivalent of the Rubin Observatory's single exposures. This is an enticing opportunity to extract vastly more detections from ecliptic plane surveys (e.g. Pan-STARRS) and synoptic surveys like ZTF (Fig.~\ref{fig:Vmagdepth}).
    \item However, doing so requires truly enormous computation that scales rapidly as the distance to the targeted populations decreases (Fig.~\ref{fig:integrateddensity}). Nonetheless, the distant solar system, including most of Planet~9's untested parameter space, can be probed with existing ZTF survey data (Sec.~\ref{sec:ztfdensity}).
    \item There are several tradeoffs that can enable tractable searches across wide portions of parameter space. We highlighted a few, such as hierarchical searches over short time spans and smoothing or binning pixels (see Sec.~\ref{sec:tradeoffs}).
    \item Targeted searches within all-sky surveys, such as the LSST deep drilling fields, dense single or few-night campaigns, or searches in restricted portions of parameter space, are feasible (though often still requiring world-class HPC) and offer avenues for high-impact research, particularly for outer solar system population studies.
\end{itemize}

\begin{acknowledgements}
The authors thank Peter McGill, Sage Li, Lila Braff, Bob Armstrong, Scott Perkins, and Michael Schneider for their input over the course of this study.

This work was performed under the auspices of the U.S. Department of Energy by Lawrence Livermore National Laboratory under Contract DE-AC52-07NA27344. This work was supported by the Lawrence Livermore National Laboratory LDRD Program under Project 2023-ERD-044. The document release number is LLNL-JRNL-2011617.

Based on observations obtained with the Samuel Oschin Telescope 48-inch and the 60-inch Telescope at the Palomar
Observatory as part of the Zwicky Transient Facility project. ZTF is supported by the National Science Foundation under Grants
No. AST-1440341 and AST-2034437 and a collaboration including current partners Caltech, IPAC, the Oskar Klein Center at
Stockholm University, the University of Maryland, University of California, Berkeley, the University of Wisconsin at Milwaukee,
University of Warwick, Ruhr University, Cornell University, Northwestern University and Drexel University. Operations are
conducted by COO, IPAC, and UW.
\end{acknowledgements}

\software{
Astropy \citep{astropy:2013, astropy:2018, astropy:2022},
Astroquery \citep{2019AJ....157...98G},
Matplotlib \citep{Hunter:2007},
numdifftools \citep{numdifftools,AdaptiveRobustNumericalDifferentiation},
NumPy \citep{2020NumPy-Array},
pandas \citep{mckinney-proc-scipy-2010,reback2020pandas},
SciPy \citep{2020SciPy-NMeth},
SSAPy \citep{ssapy1}.}


\appendix

\section{Expected detection significance in matched filtering\label{sec:app_expectedSNR}}

In this appendix we derive the expected detection significance for optimal image stacking when there is a mismatch between the model used for the matched filter and the true source. 

\subsection{Pixel sums as convolutions}

The expected detection significance using an arbitrary set of pixel weights is given by Eq.~\ref{eqn:SNR_weights},
\begin{equation}
\mathrm{SNR} = \frac{\sum_i w_i s_i}{\sqrt{\sum_i w_i^2 \sigma_i^2}}, \label{eqn:SNR_weights_app}
\end{equation}
where $s_i$ is the object's true flux in pixel $i$, $w_i$ is the weight assigned, and $\sigma_i^2$ is the noise variance in the pixel (noise in different pixels is assumed to be independent; $\sigma_i$ has the same units as $s_i$, e.g. ADU). Each sum is over all images and over the pixels within each image. For now consider the inner sum, so that the index $i$ just runs over the pixels in a single image.

The true flux $s_i$ in pixel $i$ can be written in terms of a continuous, normalized flux density $f(x)$. The flux density is scaled by the object's brightness $A$, shifted to its location within the image $x_0$, and then integrated over pixels. Mathematically,
\begin{align}
s_i &= \int\limits_\text{pixel $i$} Af(x-x_0) \, d^2 x \nonumber\\
&= \int\limits_{\mathbb{R}^2} Af(x-x_0) \, \pixarea W_\pixsize(x-x_i)\, d^2 x \nonumber\\
&= \left(Af \ast \pixarea W^-_\pixsize\right)(x_i-x_0), \label{eqn:true_flux_continuous}
\end{align}
where $x_i$ is the location of the center of pixel $i$, $\pixsize$ is the pixel side length, and we introduced the square top hat pixel density $W_\pixsize(x)$, which equals $1/\pixarea$ when $x$ is in the range $-\pixsize/2$ to $\pixsize/2$ in both dimensions and equals 0 otherwise. In the last line, the symbol $\ast$ represents 2-d convolution and the minus sign superscript denotes the mirror image of a function, i.e. $W_\pixsize^-(x) = W_\pixsize(-x)$, though in this case $W_\pixsize^- = W_\pixsize$ since the top hat function is symmetric.

The optimal weight assigned to pixel $i$ is $w_i = \ts_i/ \sigma_i^2$ (Eq.~\ref{eqn:optweights}), where $\ts_i$ is the predicted flux according to the model used for the matched filter (quantities with tildes are associated with the trial model used for the weights). Similar to $s_i$ above,
\begin{align}
w_i &= \frac{1}{\sigma^2}  \int\limits_\text{pixel $i$} \tA \tf(x-\tx_0) \, d^2 x, \label{eqn:weight_intpixel}\\
&\approx \frac{1}{\sigma^2} \tA \tf(x_i-\tx_0) \pixarea, \label{eqn:weight_pixcenter}
\end{align}
where $\tf(x-\tx_0)$ is the normalized flux density of the trial object located at position $\tx_0$ in the image. In the second line we use a modified weight that slightly trades optimality for speed of computation by approximating the integral over pixel $i$. In practice these modified weights remain very close to optimal for the PSFs and pixel scales of typical imaging systems\footnote{We numerically checked the reduction in expected significance using these suboptimal weights when $f$ and $\tf$ are 2-d Gaussian densities. The loss is negligible when the pixel scale is small compared to the PSF width. For an extreme case where the PSF full width at half max (FWHM) is 0.9 pixels, SNR decreases by 3\% when using weights from Eq.~\ref{eqn:weight_pixcenter} instead of Eq.~\ref{eqn:weight_intpixel}. For the more realistic case of a FWHM of 2 pixels the loss in SNR is less than 0.2\%.}. We also assume that the noise level is constant in the vicinity of the source so that there is a single value $\sigma^2$ for all relevant pixels in the image.

We can find closed form expressions for the inner pixel sums in Eq.~\ref{eqn:SNR_weights_app} by taking averages with respect to the location of the source within a pixel, i.e. the true position of the source is equally likely to fall anywhere relative to the pixel boundaries. This averaging is highly apposite since when searching for unknown objects, source positions within the images are completely independent from the pixel alignment. Stacking many images on a target that moves over time is a further averaging of source position relative to the pixel boundaries.

To do the averaging, first fix the separation between the trial position $\tx_0$ and true position $x_0$,
\begin{equation}
\Delta x \equiv \tx_0 - x_0 = \text{constant}, \label{eqn:Deltaxdef}
\end{equation}
and then average over the subpixel location $x_0$. Concretely, the expected value of the arbitrary function $F(x_0)$ with respect to the subpixel position of $x_0$ is
\begin{equation}
\mathrm{E}_{x_0}\left[ F(x_0) \right] = \int\limits_\text{pixel $0$} F(x_0) \frac{d^2 x_0}{\pixarea},
\end{equation}
where the integral is over the ``central pixel'', i.e. the integration limits for $x_0$ are from $-\pixsize/2$ to $+\pixsize/2$ in both dimensions. There is no loss of generality in considering the source to be located in the central pixel as opposed to an arbitrary pixel, or even randomly located over the entire plane.

From Eqs.~\ref{eqn:true_flux_continuous} and~\ref{eqn:weight_pixcenter}, the numerator of Eq.~\ref{eqn:SNR_weights_app} has the form $\sum_i g(x_i-x_0) h(x_i-\tx_0)$. Averaging over the subpixel location $x_0$ gives
\begin{align}
\mathrm{E}_{x_0} \left[ \sum_i g(x_i-x_0) h(x_i-\tx_0) \right] &= \frac{1}{\pixarea} \int\limits_\text{pixel $0$} d^2 x_0 \left(\sum_i g(x_i-x_0) h(x_i- x_0 -\Delta x) \right) \nonumber\\
&= \frac{1}{\pixarea} \sum_i \int\limits_\text{pixel $i$} d^2 x' \,  g(x') h(x' - \Delta x) \nonumber\\
&= \frac{1}{\pixarea} \int\limits_{\mathbb{R}^2} d^2 x' \, g(x') h(x'-\Delta x) \nonumber\\
&= \frac{1}{\pixarea} \left(g \ast h^- \right)(\Delta x),
\end{align}
where in the second line we exchanged the sum and integral and then changed variables to $x' = x_i - x_0$.

Using Eqs.~\ref{eqn:true_flux_continuous} and~\ref{eqn:weight_pixcenter} for $g$ and $h$, we have
\begin{equation}
\mathrm{E}_{x_0} \left[ \sum_i w_i s_i \right] = \frac{A\tA}{\sigma^2 / \pixarea} \left(f \ast \tf^- \ast W_\pixsize^- \right) (\Delta x). \label{eqn:pixsum_numerator}
\end{equation}
Similarly, the sum in the denominator of Eq.~\ref{eqn:SNR_weights_app} is
\begin{equation}
\mathrm{E}_{x_0} \left[ \sum_i w_i^2 \sigma^2 \right] = \frac{\tA^2}{\sigma^2 / \pixarea} \left(\tf \ast \tf^- \right)(0). \label{eqn:pixsum_denominator}
\end{equation}

So far we have only calculated sums over the pixels within a single image. In Eq.~\ref{eqn:SNR_weights_app} there is an additional outer sum over all the images which contain the source. The numerator and denominator in the expected significance will therefore consist of sums of many terms. We can invoke the law of large numbers to conclude that these outer sums will be close to their means, converging with increasing number of stacked images. In other words, we can use Eqs.~\ref{eqn:pixsum_numerator} and~\ref{eqn:pixsum_denominator} to compute the expected detection significance and introduce little error compared to what we would have obtained using Eq.~\ref{eqn:SNR_weights_app} without any averaging over subpixel location.

\subsection{Gaussian intensity profile}
Most solar system bodies will appear as point sources in images. In fact, the original motivation for digital tracking was to use short exposure times to prevent so-called trailing loss where a moving object's flux is spread across multiple pixels in a streak.

The normalized flux density $f(x)$ in this case simply traces the telescope point spread function (PSF) $\phi_b$, and we set $f(x) = \tf(x) = \phi_b(x)$. We model the PSF as a 2-d Gaussian with standard deviation $b$,
\begin{equation}
\phi_b(x) = \frac{1}{2\pi b^2} \exp\left( -\frac{|x|^2}{2 b^2} \right). \label{eqn:2dgaussianpdf}
\end{equation}

We do not consider streak profiles for $f(x)$ but it would be relatively straightforward to do so (perhaps at the expense of closed form expressions).

\subsection{Expected significance for small pixels}
Equations~\ref{eqn:pixsum_numerator} and~\ref{eqn:pixsum_denominator} involve the convolution of the Gaussian PSF $\phi_b$ with itself. This is simply another Gaussian with twice the variance: $\phi_b \ast \phi_b^- = \phi_{\sqrt{2}b}$. The doubled variance exactly corresponds to the standard practice of convolving an image with its PSF to create a significance map, which means a point source is convolved with the PSF twice, first by the atmosphere and then in software.

Convolution with the pixel density $W^-_\pixsize$ causes a further blur by the pixel size. If the pixels are much smaller than the PSF width ($\pixsize \ll b$) this blurring is negligible compared to the width of  $\phi_{\sqrt{2}b}$ and we can ignore the convolution with $W^-_\pixsize$:
\begin{align}
\mathrm{E}_{x_0} \left[ \sum\limits_i w_i s_i \right] &=  \frac{A\tA}{\sigma^2 / \pixarea} \phi_{\sqrt{2}b} (\Delta x), \label{eqn:pixsum_numerator_small} \\
\mathrm{E}_{x_0} \left[ \sum\limits_i w_i^2 \sigma^2 \right] &= \frac{\tA^2}{\sigma^2 / \pixarea} \phi_{\sqrt{2}b}(0).\label{eqn:pixsum_denominator_small}
\end{align}

Plugging these into Eq.~\ref{eqn:SNR_weights_app} and restoring the outer sums over all the images which contain the source we obtain
\begin{align}
\SNR(\Delta x) &= \dfrac{\displaystyle\sum\limits_i \dfrac{ \tA_i A_i}{\sigma_i^2 /\pixareai} \, \phi_i(\Delta x^i)}
{\displaystyle\sqrt{\sum\limits_i \dfrac{\tA_i^2}{\sigma_i^2 / \pixareai} \, \phi_i(0)}}. \label{eqn:SNRsmallpix_allfactors}
\end{align}

The index $i$ now runs over all images so that $b_i$, $\pixsize_i$, and $\sigma^2_i$, are the PSF width, pixel scale, and noise variance for image $i$. The amplitudes $\tA_i$ and $A_i$ are the fluxes of the trial and true source, and $\Delta x^i$ is the separation between the trial and true source positions in image $i$. For clarity we use $\phi_i$ to stand for $\phi_{\sqrt{2}b_i}$.

Note that $\SNR$ is not affected by the overall scaling of $\tA_i$ (see the discussion after Eq.~\ref{eqn:optweights}). What is important is the relative flux of the source between different images, i.e. the light curve. For our purposes, minor planet light curves are determined by the Sun-object-Earth geometry of their orbit. We will always consider trial orbits that are quite close to the true orbit. Therefore, the trial light curve is proportional to the true light curve ($\tA_i \propto A_i$), which we now use to further simplify Eq.~\ref{eqn:SNRsmallpix_allfactors}.

The maximum detection significance occurs when the predicted position of the source coincides with its true position in all images ($\Delta x^i=0$ for all~$i$). We can factor out this maximum significance to rewrite Eq.~\ref{eqn:SNRsmallpix_allfactors} as
\begin{align}
\SNR(\Delta x) &= \SNRmax \sum\limits_i c_i \frac{\phi_i (\Delta x^i)}{\phi_i(0)} \label{eqn:SNRsmallpix_compactphi} \\
&= \SNRmax \sum\limits_i c_i \exp\left(-\frac{\lvert \Delta x^i \rvert^2}{4 b_i^2}\right), \label{eqn:SNRsmallpix_compact}
\end{align}
where
\begin{align}
\SNRmax = \sqrt{ \sum\limits_i \frac{A_i^2}{4\pi \sigma_i^2 (b_i/\pixsize_i)^2} }, \label{eqn:SNRmaxsmallpix}
\end{align}
and the normalized image weights $c_i$ are defined by
\begin{equation}
c_i = \frac{c_i'}{\sum_i c_i'} \quad \text{with} \quad c_i' = \frac{\tA_i^2}{\sigma_i^2 (b_i/\pixsize_i)^2}. \label{eqn:SNRc_smallpix}
\end{equation}

Equation~\ref{eqn:SNRmaxsmallpix} is well known in astronomy as the inverse of the relative uncertainty of source's flux in background-dominated imaging \citep[e.g.][]{1983PASP...95..163K}. It is a convenient formula that can be used to find, for example, the completeness of a stacking search. One simply scales the $A_i$'s until $\SNRmax$ equals, say, 10 and then converts the scaling into a magnitude.

\subsection{Expected significance for finite pixels\label{sec:app_expectedSNR_finite}}
Most astronomical imaging is designed so that the PSF is comparable to the pixel scale. In this case, to retain full accuracy, we should include the convolution with $W^-_\pixsize$ in Eq.~\ref{eqn:pixsum_numerator}. The convolution of a Gaussian with the top hat pixel density is just the average of the Gaussian density over a pixel, which we call $\phibar_b(x)$,
\begin{equation}
\phibar_b(x) \equiv \left(\phi_b \ast W^-_\pixsize\right)(x) =  \frac{1}{\pixarea}\int\limits_\text{pixel $0$} d^2 x' \phi_{b} \left(x - x'\right), \label{eqn:avggaussian}
\end{equation}
where the integration is from $-\pixsize/2$ to $\pixsize/2$ in both the $x$ and $y$ directions (the dependence on pixel size $\pixsize$ is implicit in $\phibar_b$).

For finite size pixels Eq.~\ref{eqn:pixsum_numerator} is
\begin{equation}
\mathrm{E}_{x_0} \left[ \sum\limits_i w_i s_i \right] =  \frac{A\tA}{\sigma^2 / \pixarea} \phibar_{\sqrt{2}b} (\Delta x), \label{eqn:pixsum_numerator_finite}
\end{equation}
which is just Eq.~\ref{eqn:pixsum_numerator_small} averaged over a pixel. Equation~\ref{eqn:pixsum_denominator_small} is unchanged.

We therefore obtain slight modifications of Eqs.~\ref{eqn:SNRsmallpix_compactphi},~\ref{eqn:SNRmaxsmallpix}, and~\ref{eqn:SNRc_smallpix},
\begin{align}
\SNR(\Delta x) &= \SNRmax \sum\limits_i c_i \frac{\phibar_i(\Delta x^i)}{\phibar_i(0)}, \label{eqn:SNRexact_app} \\
\SNRmax &= \dfrac{\displaystyle\sum\limits_i \dfrac{A_i^2 \qzero_i}{4\pi \sigma_i^2 (b_i/\pixsize_i)^2}}
{\displaystyle\sqrt{\sum\limits_i \dfrac{A_i^2}{4\pi \sigma_i^2 (b_i/\pixsize_i)^2}}}, \label{eqn:SNRmaxexact_app} \\
\quad c_i = \frac{c'_i}{\sum_i c'_i}, \quad &\text{where}\quad c'_i = \frac{\tA_i^2 \qzero_i}{\sigma_i^2 (b_i/\pixsize_i)^2}. \label{eqn:SNRc_exact_app}
\end{align}
In these equations $\phibar_i$ is shorthand for $\phibar_{\sqrt{2}b_i}$ and $\qzero_i$ is shorthand for $\qzero_{\pixsize_i/\sqrt{2}b_i}$ (defined in Eq.~\ref{eqn:finitepixcorrection0def}). The latter is a pixel size correction factor that converges to 1 for small pixels, meaning that Eqs.~\ref{eqn:SNRexact_app},~\ref{eqn:SNRmaxexact_app}, and~\ref{eqn:SNRc_exact_app} reduce to Eqs.~\ref{eqn:SNRsmallpix_compactphi},~\ref{eqn:SNRmaxsmallpix}, and~\ref{eqn:SNRc_smallpix} in that limit. 

\section{Properties of \lowercase{$\phibar_b(x)$}\label{sec:phibar}}
In Sec.~\ref{sec:exactmetric} we use the Taylor series of the average 2-d Gaussian density $\phibar_b(x)$ around $x=0$. Since $\phibar_b(x)$ is an even function of $x$, the series only contains even powers of $x$. The lowest order terms are
\begin{align}
\phibar_b(x)  &= \phibar_b(0) - \frac{1}{2} \sqrt{\phibar_b(0)} \, \frac{\exp\left(-\frac{(\pixsize/2)^2}{2b^2}\right)}{\sqrt{2\pi b^2}} \left\lvert\frac{x}{b}\right\rvert^2 + \mathcal{O}\left\lvert\frac{x}{b}\right\rvert^4 \nonumber.
\end{align}
We introduce two pixel size correction factors $\qzero_{\pixsize/b}$ and $\qtwo_{\pixsize/b}$ to isolate the effects of the finite pixel size. First,
\begin{align}
\phibar_b(0) \equiv \phi_b(0) \, \qzero_{\pixsize/b}, \label{eqn:avggaussianatzero}
\end{align}
where $\phi_b(0)=1/(2\pi b^2)$ is the $\pixsize \to 0$ limit so that
\begin{equation}
\qzero_{\pixsize/b} = \left[\frac{\sqrt{2\pi}}{\pixsize/b} \erf\left(\frac{\pixsize/b}{2\sqrt{2}}\right)\right]^2, \label{eqn:finitepixcorrection0def}
\end{equation}
which converges to 1 when $\pixsize \lesssim b$ (i.e. when the pixel size is negligible).

Next, we define
\begin{equation}
\qtwo_{\pixsize/b} \equiv \frac{\exp\left(-\frac{1}{8}(\pixsize/b)^2\right)}{\sqrt{\qzero_{\pixsize/b}}} 
= \frac{\exp\left(-\frac{1}{8}(\pixsize/b)^2\right)}{\frac{\sqrt{2\pi}}{\pixsize/b} \erf\left(\frac{\pixsize/b}{2\sqrt{2}}\right)}, \label{eqn:finitepixcorrection2def}
\end{equation}
where both the numerator and denominator are $\approx 1$ when $\pixsize \lesssim b$.

Using Eq.~\ref{eqn:finitepixcorrection2def}, the Taylor series for $\phibar_b(x)$ can be written
\begin{align}
\frac{\phibar_b(x)}{\phibar_b(0)} = 1 - \frac{1}{2} \qtwo_{\pixsize/b} \left\lvert\frac{x}{b}\right\rvert^2 + \mathcal{O}\left\lvert\frac{x}{b}\right\rvert^4 \label{eqn:avggaussianseries},
\end{align}
which is needed in Sec.~\ref{sec:exactmetric}.

\begin{figure}
\begin{center}
\includegraphics{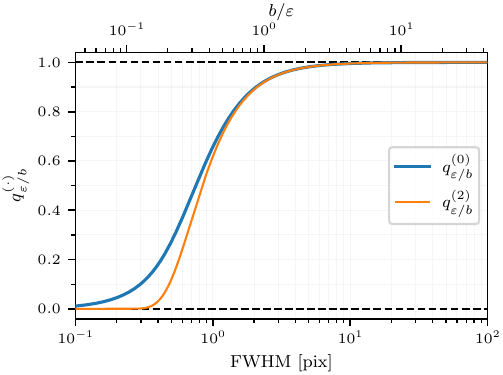}
\end{center}
\caption{\label{fig:pixelfactors} The pixel size correction factors $\qzero_{\pixsize/b}$ (Eq.~\ref{eqn:finitepixcorrection0def}) and $\qtwo_{\pixsize/b}$ (Eq.~\ref{eqn:finitepixcorrection2def}) for a range of PSF sizes.}
\end{figure}

Figure~\ref{fig:pixelfactors} shows the two pixel size correction factors as a function of $b/\pixsize$ and the PSF full width at half max ($\mathrm{FWHM} = \left(2\sqrt{2 \log 2}\right) b$ for a Gaussian PSF). The correction factors are $\gtrsim 0.9$ when the $\mathrm{FWHM} \gtrsim 2~\mathrm{pixels}$ but can be significantly less than one for narrower PSFs.

\section{Feasibility of Eight-Hour Continuous Tracking with Airmass $\leq 2$ at Cerro Pach\'on\label{sec:continuousviewingfield}}

This appendix derives the condition under which a fixed field can be tracked from the ground for at least $8$~hours while remaining at airmass $\leq 2$. Refraction and local horizon obstructions are neglected; using a more accurate airmass model only shifts the altitude threshold slightly and does not change the algebra below.

Altitude $h$ is elevation above the horizon. The zenith angle is $z=90^\circ-h$. The hour angle $H$, local sidereal time, and right ascension of the field $\alpha$ are related by $H=\mathrm{LST}-\alpha$ ($H$ is zero at upper transit and increases westward at $15^\circ$ per hour). The declination of the field is $\delta$. Assuming the airmass is $X\approx \sec z$, the requirement $X\le 2$ gives $z\le 60^\circ$, i.e. the elevation must be $h \ge h_0 \equiv 30^\circ$.

\noindent The standard spherical relation is
\begin{equation*}
\sin h = \sin\phi\,\sin\delta + \cos\phi\,\cos\delta\,\cos H,
\end{equation*}
where $\phi$ is the latitude of the observatory. The airmass requirement is $\sin h \ge \sin h_0$, which implies
\begin{equation}
    \cos H \;\ge\; C(\delta)
    \;\equiv\;
    \frac{\sin h_0-\sin\phi\,\sin\delta}{\cos\phi\,\cos\delta}.
    \label{eqn:Cdelta}
\end{equation}
If $\vert C(\delta) \vert\le 1$, the allowed hour angles are $|H|\le H_0$ where $H_0 \equiv \arccos\!\big(C(\delta)\big)$. The continuous observation period meeting the airmass requirement is therefore
\begin{equation*}
    t(\delta;h_0,\phi)
    =\frac{2H_0}{15^\circ/\mathrm{hr}}.
    \label{eq:time}
\end{equation*}

\noindent Requiring $t\ge 8$~hours gives $H_0\ge 60^\circ$, i.e.\ $\cos H_0 = C(\delta) \le\tfrac{1}{2}$. Using the definition of $C(\delta)$ (Eq.~\ref{eqn:Cdelta}), this yields a constraint on the declination of the field,
\begin{equation*}
\sin\phi\,\sin\delta + \tfrac{1}{2}\cos\phi\,\cos\delta \ge \sin h_0.
\end{equation*}

\noindent Inserting the latitude of the Rubin Observatory at Cerro Pach\'on ($\phi=-30.24^\circ$) and solving for the allowed declination gives
\begin{equation*}
\delta \le -8.3^\circ.
\end{equation*}
This means that a field at $\delta = -8.3^\circ$ is visible at airmass less than 2 for 8~hours; fields further south can be viewed for longer.

If we are interested in observing along the ecliptic plane, we can solve for the viable ecliptic longitudes $\lambda$ using $\sin\left(\delta(\lambda)\right) = \sin\varepsilon\,\sin\lambda$, where $\varepsilon=23.44^\circ$ is Earth's obliquity. The result is
\begin{equation*}
    \lambda \in [\,201.2^\circ,\,338.8^\circ\,].
\end{equation*}

\bibliography{mybibfile}{}

\begin{thebibliography}{}
\expandafter\ifx\csname natexlab\endcsname\relax\def\natexlab#1{#1}\fi
\providecommand{\url}[1]{\href{#1}{#1}}
\providecommand{\dodoi}[1]{doi:~\href{http://doi.org/#1}{\nolinkurl{#1}}}
\providecommand{\doeprint}[1]{\href{http://ascl.net/#1}{\nolinkurl{http://ascl.net/#1}}}
\providecommand{\doarXiv}[1]{\href{https://arxiv.org/abs/#1}{\nolinkurl{https://arxiv.org/abs/#1}}}

\bibitem[{R.~L. {Allen} {et~al.}(2001){Allen}, {Bernstein}, \&
  {Malhotra}}]{2001ApJ...549L.241A}
{Allen}, R.~L., {Bernstein}, G.~M., \& {Malhotra}, R. 2001,
  \bibinfo{title}{{The Edge of the Solar System},} \apjl, 549, L241,
  \dodoi{10.1086/319165}

\bibitem[{A. {Alvarez-Candal} {et~al.}(2020){Alvarez-Candal},
  {Souza-Feliciano}, {Martins-Filho}, {Pinilla-Alonso}, \&
  {Ortiz}}]{2020MNRAS.497.5473A}
{Alvarez-Candal}, A., {Souza-Feliciano}, A.~C., {Martins-Filho}, W.,
  {Pinilla-Alonso}, N., \& {Ortiz}, J.~L. 2020, \bibinfo{title}{{The dwarf
  planet Makemake as seen by X-Shooter},} \mnras, 497, 5473,
  \dodoi{10.1093/mnras/staa2329}

\bibitem[{S.-i. {Amari}(1985){Amari}}]{amari1985differential}
{Amari}, S.-i. 1985, Lecture Notes in Statistics, Vol.~28,
  Differential-Geometrical Methods in Statistics, 1st edn. (New York, NY:
  Springer), \dodoi{10.1007/978-1-4612-5056-2}

\bibitem[{ {ASTM International}(2019){ASTM International}}]{ASTM_E490-00AR19}
{ASTM International}. 2019, Standard Solar Constant and Zero Air Mass Solar
  Spectral Irradiance Tables, {ASTM International},
  \dodoi{10.1520/E0490-00AR19}

\bibitem[{ {Astropy Collaboration} {et~al.}(2013){Astropy Collaboration},
  {Robitaille}, {Tollerud}, {Greenfield}, {Droettboom}, {Bray}, {Aldcroft},
  {Davis}, {Ginsburg}, {Price-Whelan}, {Kerzendorf}, {Conley}, {Crighton},
  {Barbary}, {Muna}, {Ferguson}, {Grollier}, {Parikh}, {Nair}, {Unther},
  {Deil}, {Woillez}, {Conseil}, {Kramer}, {Turner}, {Singer}, {Fox}, {Weaver},
  {Zabalza}, {Edwards}, {Azalee Bostroem}, {Burke}, {Casey}, {Crawford},
  {Dencheva}, {Ely}, {Jenness}, {Labrie}, {Lim}, {Pierfederici}, {Pontzen},
  {Ptak}, {Refsdal}, {Servillat}, \& {Streicher}}]{astropy:2013}
{Astropy Collaboration}, {Robitaille}, T.~P., {Tollerud}, E.~J., {et~al.} 2013,
  \bibinfo{title}{{Astropy: A community Python package for astronomy},} \aap,
  558, A33, \dodoi{10.1051/0004-6361/201322068}

\bibitem[{ {Astropy Collaboration} {et~al.}(2018){Astropy Collaboration},
  {Price-Whelan}, {Sip{\H{o}}cz}, {G{\"u}nther}, {Lim}, {Crawford}, {Conseil},
  {Shupe}, {Craig}, {Dencheva}, {Ginsburg}, {Vand erPlas}, {Bradley},
  {P{\'e}rez-Su{\'a}rez}, {de Val-Borro}, {Aldcroft}, {Cruz}, {Robitaille},
  {Tollerud}, {Ardelean}, {Babej}, {Bach}, {Bachetti}, {Bakanov}, {Bamford},
  {Barentsen}, {Barmby}, {Baumbach}, {Berry}, {Biscani}, {Boquien}, {Bostroem},
  {Bouma}, {Brammer}, {Bray}, {Breytenbach}, {Buddelmeijer}, {Burke},
  {Calderone}, {Cano Rodr{\'\i}guez}, {Cara}, {Cardoso}, {Cheedella}, {Copin},
  {Corrales}, {Crichton}, {D'Avella}, {Deil}, {Depagne}, {Dietrich}, {Donath},
  {Droettboom}, {Earl}, {Erben}, {Fabbro}, {Ferreira}, {Finethy}, {Fox},
  {Garrison}, {Gibbons}, {Goldstein}, {Gommers}, {Greco}, {Greenfield},
  {Groener}, {Grollier}, {Hagen}, {Hirst}, {Homeier}, {Horton}, {Hosseinzadeh},
  {Hu}, {Hunkeler}, {Ivezi{\'c}}, {Jain}, {Jenness}, {Kanarek}, {Kendrew},
  {Kern}, {Kerzendorf}, {Khvalko}, {King}, {Kirkby}, {Kulkarni}, {Kumar},
  {Lee}, {Lenz}, {Littlefair}, {Ma}, {Macleod}, {Mastropietro}, {McCully},
  {Montagnac}, {Morris}, {Mueller}, {Mumford}, {Muna}, {Murphy}, {Nelson},
  {Nguyen}, {Ninan}, {N{\"o}the}, {Ogaz}, {Oh}, {Parejko}, {Parley}, {Pascual},
  {Patil}, {Patil}, {Plunkett}, {Prochaska}, {Rastogi}, {Reddy Janga},
  {Sabater}, {Sakurikar}, {Seifert}, {Sherbert}, {Sherwood-Taylor}, {Shih},
  {Sick}, {Silbiger}, {Singanamalla}, {Singer}, {Sladen}, {Sooley},
  {Sornarajah}, {Streicher}, {Teuben}, {Thomas}, {Tremblay}, {Turner},
  {Terr{\'o}n}, {van Kerkwijk}, {de la Vega}, {Watkins}, {Weaver}, {Whitmore},
  {Woillez}, {Zabalza}, \& {Astropy Contributors}}]{astropy:2018}
{Astropy Collaboration}, {Price-Whelan}, A.~M., {Sip{\H{o}}cz}, B.~M., {et~al.}
  2018, \bibinfo{title}{{The Astropy Project: Building an Open-science Project
  and Status of the v2.0 Core Package},} \aj, 156, 123,
  \dodoi{10.3847/1538-3881/aabc4f}

\bibitem[{ {Astropy Collaboration} {et~al.}(2022){Astropy Collaboration},
  {Price-Whelan}, {Lim}, {Earl}, {Starkman}, {Bradley}, {Shupe}, {Patil},
  {Corrales}, {Brasseur}, {N{"o}the}, {Donath}, {Tollerud}, {Morris},
  {Ginsburg}, {Vaher}, {Weaver}, {Tocknell}, {Jamieson}, {van Kerkwijk},
  {Robitaille}, {Merry}, {Bachetti}, {G{"u}nther}, {Aldcroft},
  {Alvarado-Montes}, {Archibald}, {B{'o}di}, {Bapat}, {Barentsen}, {Baz{'a}n},
  {Biswas}, {Boquien}, {Burke}, {Cara}, {Cara}, {Conroy}, {Conseil}, {Craig},
  {Cross}, {Cruz}, {D'Eugenio}, {Dencheva}, {Devillepoix}, {Dietrich},
  {Eigenbrot}, {Erben}, {Ferreira}, {Foreman-Mackey}, {Fox}, {Freij}, {Garg},
  {Geda}, {Glattly}, {Gondhalekar}, {Gordon}, {Grant}, {Greenfield}, {Groener},
  {Guest}, {Gurovich}, {Handberg}, {Hart}, {Hatfield-Dodds}, {Homeier},
  {Hosseinzadeh}, {Jenness}, {Jones}, {Joseph}, {Kalmbach}, {Karamehmetoglu},
  {Ka{l}uszy{'n}ski}, {Kelley}, {Kern}, {Kerzendorf}, {Koch}, {Kulumani},
  {Lee}, {Ly}, {Ma}, {MacBride}, {Maljaars}, {Muna}, {Murphy}, {Norman},
  {O'Steen}, {Oman}, {Pacifici}, {Pascual}, {Pascual-Granado}, {Patil},
  {Perren}, {Pickering}, {Rastogi}, {Roulston}, {Ryan}, {Rykoff}, {Sabater},
  {Sakurikar}, {Salgado}, {Sanghi}, {Saunders}, {Savchenko}, {Schwardt},
  {Seifert-Eckert}, {Shih}, {Jain}, {Shukla}, {Sick}, {Simpson},
  {Singanamalla}, {Singer}, {Singhal}, {Sinha}, {Sip{H{o}}cz}, {Spitler},
  {Stansby}, {Streicher}, {{{S}}umak}, {Swinbank}, {Taranu}, {Tewary},
  {Tremblay}, {Val-Borro}, {Van Kooten}, {Vasovi{'c}}, {Verma}, {de Miranda
  Cardoso}, {Williams}, {Wilson}, {Winkel}, {Wood-Vasey}, {Xue}, {Yoachim},
  {Zhang}, {Zonca}, \& {Astropy Project Contributors}}]{astropy:2022}
{Astropy Collaboration}, {Price-Whelan}, A.~M., {Lim}, P.~L., {et~al.} 2022,
  \bibinfo{title}{{The Astropy Project: Sustaining and Growing a
  Community-oriented Open-source Project and the Latest Major Release (v5.0) of
  the Core Package},} \apj, 935, 167, \dodoi{10.3847/1538-4357/ac7c74}

\bibitem[{K. {Batygin} {et~al.}(2019){Batygin}, {Adams}, {Brown}, \&
  {Becker}}]{2019PhR...805....1B}
{Batygin}, K., {Adams}, F.~C., {Brown}, M.~E., \& {Becker}, J.~C. 2019,
  \bibinfo{title}{{The planet nine hypothesis},} \physrep, 805, 1,
  \dodoi{10.1016/j.physrep.2019.01.009}

\bibitem[{K. {Batygin} \& M.~E. {Brown}(2016){Batygin} \&
  {Brown}}]{2016AJ....151...22B}
{Batygin}, K., \& {Brown}, M.~E. 2016, \bibinfo{title}{{Evidence for a Distant
  Giant Planet in the Solar System},} \aj, 151, 22,
  \dodoi{10.3847/0004-6256/151/2/22}

\bibitem[{E.~C. {Bellm} {et~al.}(2019{\natexlab{a}}){Bellm}, {Kulkarni},
  {Graham}, {Dekany}, {Smith}, {Riddle}, {Masci}, {Helou}, {Prince}, {Adams},
  {Barbarino}, {Barlow}, {Bauer}, {Beck}, {Belicki}, {Biswas}, {Blagorodnova},
  {Bodewits}, {Bolin}, {Brinnel}, {Brooke}, {Bue}, {Bulla}, {Burruss}, {Cenko},
  {Chang}, {Connolly}, {Coughlin}, {Cromer}, {Cunningham}, {De}, {Delacroix},
  {Desai}, {Duev}, {Eadie}, {Farnham}, {Feeney}, {Feindt}, {Flynn},
  {Franckowiak}, {Frederick}, {Fremling}, {Gal-Yam}, {Gezari}, {Giomi},
  {Goldstein}, {Golkhou}, {Goobar}, {Groom}, {Hacopians}, {Hale}, {Henning},
  {Ho}, {Hover}, {Howell}, {Hung}, {Huppenkothen}, {Imel}, {Ip}, {Ivezi{\'c}},
  {Jackson}, {Jones}, {Juric}, {Kasliwal}, {Kaspi}, {Kaye}, {Kelley},
  {Kowalski}, {Kramer}, {Kupfer}, {Landry}, {Laher}, {Lee}, {Lin}, {Lin},
  {Lunnan}, {Giomi}, {Mahabal}, {Mao}, {Miller}, {Monkewitz}, {Murphy},
  {Ngeow}, {Nordin}, {Nugent}, {Ofek}, {Patterson}, {Penprase}, {Porter},
  {Rauch}, {Rebbapragada}, {Reiley}, {Rigault}, {Rodriguez}, {van Roestel},
  {Rusholme}, {van Santen}, {Schulze}, {Shupe}, {Singer}, {Soumagnac}, {Stein},
  {Surace}, {Sollerman}, {Szkody}, {Taddia}, {Terek}, {Van Sistine}, {van
  Velzen}, {Vestrand}, {Walters}, {Ward}, {Ye}, {Yu}, {Yan}, \&
  {Zolkower}}]{2019PASP..131a8002B}
{Bellm}, E.~C., {Kulkarni}, S.~R., {Graham}, M.~J., {et~al.}
  2019{\natexlab{a}}, \bibinfo{title}{{The Zwicky Transient Facility: System
  Overview, Performance, and First Results},} \pasp, 131, 018002,
  \dodoi{10.1088/1538-3873/aaecbe}

\bibitem[{E.~C. {Bellm} {et~al.}(2019{\natexlab{b}}){Bellm}, {Kulkarni},
  {Barlow}, {Feindt}, {Graham}, {Goobar}, {Kupfer}, {Ngeow}, {Nugent}, {Ofek},
  {Prince}, {Riddle}, {Walters}, \& {Ye}}]{2019PASP..131f8003B}
{Bellm}, E.~C., {Kulkarni}, S.~R., {Barlow}, T., {et~al.} 2019{\natexlab{b}},
  \bibinfo{title}{{The Zwicky Transient Facility: Surveys and Scheduler},}
  \pasp, 131, 068003, \dodoi{10.1088/1538-3873/ab0c2a}

\bibitem[{M. {Belyakov} {et~al.}(2022){Belyakov}, {Bernardinelli}, \&
  {Brown}}]{2022AJ....163..216B}
{Belyakov}, M., {Bernardinelli}, P.~H., \& {Brown}, M.~E. 2022,
  \bibinfo{title}{{Limits on the Detection of Planet Nine in the Dark Energy
  Survey},} \aj, 163, 216, \dodoi{10.3847/1538-3881/ac5c56}

\bibitem[{G. {Bernstein} \& B. {Khushalani}(2000){Bernstein} \&
  {Khushalani}}]{2000AJ....120.3323B}
{Bernstein}, G., \& {Khushalani}, B. 2000, \bibinfo{title}{{Orbit Fitting and
  Uncertainties for Kuiper Belt Objects},} \aj, 120, 3323,
  \dodoi{10.1086/316868}

\bibitem[{G.~M. {Bernstein} {et~al.}(2004){Bernstein}, {Trilling}, {Allen},
  {Brown}, {Holman}, \& {Malhotra}}]{2004AJ....128.1364B}
{Bernstein}, G.~M., {Trilling}, D.~E., {Allen}, R.~L., {et~al.} 2004,
  \bibinfo{title}{{The Size Distribution of Trans-Neptunian Bodies},} \aj, 128,
  1364, \dodoi{10.1086/422919}

\bibitem[{F.~B. {Bianco} {et~al.}(2022){Bianco}, {Ivezi{\'c}}, {Jones},
  {Graham}, {Marshall}, {Saha}, {Strauss}, {Yoachim}, {Ribeiro}, {Anguita},
  {Bauer}, {Bauer}, {Bellm}, {Blum}, {Brandt}, {Brough}, {Catelan}, {Clarkson},
  {Connolly}, {Gawiser}, {Gizis}, {Hlo{\v{z}}ek}, {Kaviraj}, {Liu}, {Lochner},
  {Mahabal}, {Mandelbaum}, {McGehee}, {Neilsen}, {Olsen}, {Peiris}, {Rhodes},
  {Richards}, {Ridgway}, {Schwamb}, {Scolnic}, {Shemmer}, {Slater}, {Slosar},
  {Smartt}, {Strader}, {Street}, {Trilling}, {Verma}, {Vivas}, {Wechsler}, \&
  {Willman}}]{2022ApJS..258....1B}
{Bianco}, F.~B., {Ivezi{\'c}}, {\v{Z}}., {Jones}, R.~L., {et~al.} 2022,
  \bibinfo{title}{{Optimization of the Observing Cadence for the Rubin
  Observatory Legacy Survey of Space and Time: A Pioneering Process of
  Community-focused Experimental Design},} \apjs, 258, 1,
  \dodoi{10.3847/1538-4365/ac3e72}

\bibitem[{E. {Bowell} {et~al.}(1989){Bowell}, {Hapke}, {Domingue}, {Lumme},
  {Peltoniemi}, \& {Harris}}]{1989aste.conf..524B}
{Bowell}, E., {Hapke}, B., {Domingue}, D., {et~al.} 1989,
  \bibinfo{title}{{Application of photometric models to asteroids.},} in
  Asteroids II, ed. R.~P. {Binzel}, T.~{Gehrels}, \& M.~S. {Matthews}, 524--556

\bibitem[{L. {Brewin}(1996){Brewin}}]{brewinnormalcoordnotes}
{Brewin}, L. 1996, {Riemann Normal Coordinates}, {Department of Mathematics
  Preprint}, {Monash University}, {Clayton, Vic. 3168}.
\newblock
  \url{https://users.monash.edu.au/~leo/research/papers/files/lcb96-01.pdf}

\bibitem[{P.~A. Brodtkorb(2025)Brodtkorb}]{numdifftools}
Brodtkorb, P.~A. 2025, Numdifftools: Numerical differentiation tools, 0.9.41,
  \url{https://github.com/pbrod/numdifftools}

\bibitem[{M.~E. {Brown} \& K. {Batygin}(2021){Brown} \&
  {Batygin}}]{2021AJ....162..219B}
{Brown}, M.~E., \& {Batygin}, K. 2021, \bibinfo{title}{{The Orbit of Planet
  Nine},} \aj, 162, 219, \dodoi{10.3847/1538-3881/ac2056}

\bibitem[{M.~E. {Brown} \& K. {Batygin}(2022){Brown} \&
  {Batygin}}]{2022AJ....163..102B}
{Brown}, M.~E., \& {Batygin}, K. 2022, \bibinfo{title}{{A Search for Planet
  Nine using the Zwicky Transient Facility Public Archive},} \aj, 163, 102,
  \dodoi{10.3847/1538-3881/ac32dd}

\bibitem[{M.~E. {Brown} {et~al.}(2024){Brown}, {Holman}, \&
  {Batygin}}]{2024AJ....167..146B}
{Brown}, M.~E., {Holman}, M.~J., \& {Batygin}, K. 2024, \bibinfo{title}{{A
  Pan-STARRS1 Search for Planet Nine},} \aj, 167, 146,
  \dodoi{10.3847/1538-3881/ad24e9}

\bibitem[{A.~Y. {Burdanov} {et~al.}(2023){Burdanov}, {Hasler}, \& {de
  Wit}}]{2023MNRAS.521.4568B}
{Burdanov}, A.~Y., {Hasler}, S.~N., \& {de Wit}, J. 2023,
  \bibinfo{title}{{GPU-based framework for detecting small Solar system bodies
  in targeted exoplanet surveys},} \mnras, 521, 4568,
  \dodoi{10.1093/mnras/stad808}

\bibitem[{M.~R. {Calabretta} \& E.~W. {Greisen}(2002){Calabretta} \&
  {Greisen}}]{2002A&A...395.1077C}
{Calabretta}, M.~R., \& {Greisen}, E.~W. 2002, \bibinfo{title}{{Representations
  of celestial coordinates in FITS},} \aap, 395, 1077,
  \dodoi{10.1051/0004-6361:20021327}

\bibitem[{E.~I. {Chiang} \& M.~E. {Brown}(1999){Chiang} \&
  {Brown}}]{1999AJ....118.1411C}
{Chiang}, E.~I., \& {Brown}, M.~E. 1999, \bibinfo{title}{{Keck Pencil-Beam
  Survey for Faint Kuiper Belt Objects},} \aj, 118, 1411,
  \dodoi{10.1086/301005}

\bibitem[{A.~L. {Cochran} {et~al.}(1995){Cochran}, {Levison}, {Stern}, \&
  {Duncan}}]{1995ApJ...455..342C}
{Cochran}, A.~L., {Levison}, H.~F., {Stern}, S.~A., \& {Duncan}, M.~J. 1995,
  \bibinfo{title}{{The Discovery of Halley-sized Kuiper Belt Objects Using the
  Hubble Space Telescope},} \apj, 455, 342, \dodoi{10.1086/176581}

\bibitem[{J. D'Errico(2006)D'Errico}]{AdaptiveRobustNumericalDifferentiation}
D'Errico, J. 2006, Adaptive Robust Numerical Differentiation,,
  \url{https://www.mathworks.com/matlabcentral/fileexchange/13490-adaptive-robust-numerical-differentiation}

\bibitem[{S.~V. {Dhurandhar} \& B.~S. {Sathyaprakash}(1994){Dhurandhar} \&
  {Sathyaprakash}}]{1994PhRvD..49.1707D}
{Dhurandhar}, S.~V., \& {Sathyaprakash}, B.~S. 1994, \bibinfo{title}{{Choice of
  filters for the detection of gravitational waves from coalescing binaries.
  II. Detection in colored noise},} \prd, 49, 1707,
  \dodoi{10.1103/PhysRevD.49.1707}

\bibitem[{W.~C. {Fraser} {et~al.}(2014){Fraser}, {Brown}, {Morbidelli},
  {Parker}, \& {Batygin}}]{2014ApJ...782..100F}
{Fraser}, W.~C., {Brown}, M.~E., {Morbidelli}, A., {Parker}, A., \& {Batygin},
  K. 2014, \bibinfo{title}{{The Absolute Magnitude Distribution of Kuiper Belt
  Objects},} \apj, 782, 100, \dodoi{10.1088/0004-637X/782/2/100}

\bibitem[{W.~C. {Fraser} {et~al.}(2008){Fraser}, {Kavelaars}, {Holman},
  {Pritchet}, {Gladman}, {Grav}, {Jones}, {MacWilliams}, \&
  {Petit}}]{2008Icar..195..827F}
{Fraser}, W.~C., {Kavelaars}, J.~J., {Holman}, M.~J., {et~al.} 2008,
  \bibinfo{title}{{The Kuiper belt luminosity function from m(R)=21 to 26},}
  \icarus, 195, 827, \dodoi{10.1016/j.icarus.2008.01.014}

\bibitem[{W.~C. {Fraser} {et~al.}(2024){Fraser}, {Porter}, {Peltier},
  {Kavelaars}, {Verbiscer}, {Buie}, {Stern}, {Spencer}, {Benecchi}, {Terai},
  {Ito}, {Yoshida}, {Gerdes}, {Napier}, {Lin}, {Gwyn}, {Smotherman}, {Fabbro},
  {Singer}, {Alexander}, {Arimatsu}, {Banks}, {Bray}, {Ramy El-Maarry},
  {Ferrell}, {Fuse}, {Glass}, {Holt}, {Hong}, {Ishimaru}, {Johnson}, {Lauer},
  {Leiva}, {S. Lykawka}, {Marschall}, {N{\'u}{\~n}ez}, {Postman}, {Quirico},
  {Rhoden}, {Simpson}, {Schenk}, {Skrutskie}, {Steffl}, \&
  {Throop}}]{2024PSJ.....5..227F}
{Fraser}, W.~C., {Porter}, S.~B., {Peltier}, L., {et~al.} 2024,
  \bibinfo{title}{{Candidate Distant Trans-Neptunian Objects Detected by the
  New Horizons Subaru TNO Survey},} \psj, 5, 227, \dodoi{10.3847/PSJ/ad6f9e}

\bibitem[{C.~I. {Fuentes} {et~al.}(2009){Fuentes}, {George}, \&
  {Holman}}]{2009ApJ...696...91F}
{Fuentes}, C.~I., {George}, M.~R., \& {Holman}, M.~J. 2009, \bibinfo{title}{{A
  Subaru Pencil-Beam Search for m$_{R}$ \raisebox{-0.5ex}\textasciitilde 27
  Trans-Neptunian Bodies},} \apj, 696, 91, \dodoi{10.1088/0004-637X/696/1/91}

\bibitem[{A. {Geringer-Sameth} \& S.~M. {Koushiappas}(2012){Geringer-Sameth} \&
  {Koushiappas}}]{2012MNRAS.425..862G}
{Geringer-Sameth}, A., \& {Koushiappas}, S.~M. 2012, \bibinfo{title}{{Detecting
  unresolved moving sources in a diffuse background},} \mnras, 425, 862,
  \dodoi{10.1111/j.1365-2966.2012.21139.x}

\bibitem[{A. {Geringer-Sameth} {et~al.}(2015){Geringer-Sameth}, {Koushiappas},
  \& {Walker}}]{2015PhRvD..91h3535G}
{Geringer-Sameth}, A., {Koushiappas}, S.~M., \& {Walker}, M.~G. 2015,
  \bibinfo{title}{{Comprehensive search for dark matter annihilation in dwarf
  galaxies},} \prd, 91, 083535, \dodoi{10.1103/PhysRevD.91.083535}

\bibitem[{A. {Ginsburg} {et~al.}(2019){Ginsburg}, {Sip{\H{o}}cz}, {Brasseur},
  {Cowperthwaite}, {Craig}, {Deil}, {Guillochon}, {Guzman}, {Liedtke}, {Lian
  Lim}, {Lockhart}, {Mommert}, {Morris}, {Norman}, {Parikh}, {Persson},
  {Robitaille}, {Segovia}, {Singer}, {Tollerud}, {de Val-Borro}, {Valtchanov},
  {Woillez}, {Astroquery Collaboration}, \& {a subset of astropy
  Collaboration}}]{2019AJ....157...98G}
{Ginsburg}, A., {Sip{\H{o}}cz}, B.~M., {Brasseur}, C.~E., {et~al.} 2019,
  \bibinfo{title}{{astroquery: An Astronomical Web-querying Package in
  Python},} \aj, 157, 98, \dodoi{10.3847/1538-3881/aafc33}

\bibitem[{J.~D. {Giorgini} \&  {JPL Solar System Dynamics Group}({}){Giorgini}
  \& {JPL Solar System Dynamics Group}}]{NASAJPLHorizons}
{Giorgini}, J.~D., \& {JPL Solar System Dynamics Group}. {}, {NASA/JPL Horizons
  On-Line Ephemeris System}, \url{{https://ssd.jpl.nasa.gov/horizons}}

\bibitem[{J.~D. {Giorgini} {et~al.}(1996){Giorgini}, {Yeomans}, {Chamberlin},
  {Chodas}, {Jacobson}, {Keesey}, {Lieske}, {Ostro}, {Standish}, \&
  {Wimberly}}]{1996DPS....28.2504G}
{Giorgini}, J.~D., {Yeomans}, D.~K., {Chamberlin}, A.~B., {et~al.} 1996,
  \bibinfo{title}{{JPL's On-Line Solar System Data Service},} in AAS/Division
  for Planetary Sciences Meeting Abstracts, Vol.~28, AAS/Division for Planetary
  Sciences Meeting Abstracts \#28, 25.04

\bibitem[{B. {Gladman} {et~al.}(1998){Gladman}, {Kavelaars}, {Nicholson},
  {Loredo}, \& {Burns}}]{1998AJ....116.2042G}
{Gladman}, B., {Kavelaars}, J.~J., {Nicholson}, P.~D., {Loredo}, T.~J., \&
  {Burns}, J.~A. 1998, \bibinfo{title}{{Pencil-Beam Surveys for Faint
  Trans-Neptunian Objects},} \aj, 116, 2042, \dodoi{10.1086/300573}

\bibitem[{N. {Golovich} {et~al.}(2025){Golovich}, {Steil}, {Geringer-Sameth},
  {Iwabuchi}, {Dozier}, \& {Pearce}}]{2025A&C....5300987G}
{Golovich}, N., {Steil}, T., {Geringer-Sameth}, A., {et~al.} 2025,
  \bibinfo{title}{{Survey-wide asteroid discovery with a high-performance
  computing enabled non-linear digital tracking framework},} Astronomy and
  Computing, 53, 100987, \dodoi{10.1016/j.ascom.2025.100987}

\bibitem[{M. {Granvik} {et~al.}(2018){Granvik}, {Morbidelli}, {Jedicke},
  {Bolin}, {Bottke}, {Beshore}, {Vokrouhlick{\'y}}, {Nesvorn{\'y}}, \&
  {Michel}}]{2018Icar..312..181G}
{Granvik}, M., {Morbidelli}, A., {Jedicke}, R., {et~al.} 2018,
  \bibinfo{title}{{Debiased orbit and absolute-magnitude distributions for
  near-Earth objects},} \icarus, 312, 181, \dodoi{10.1016/j.icarus.2018.04.018}

\bibitem[{P.~S. {Gural} {et~al.}(2018){Gural}, {Otto}, \&
  {Tedesco}}]{2018PASP..130g4504G}
{Gural}, P.~S., {Otto}, P.~R., \& {Tedesco}, E.~F. 2018,
  \bibinfo{title}{{Moving Object Detection Using a Parallax Shift Vector
  Algorithm},} \pasp, 130, 074504, \dodoi{10.1088/1538-3873/aac1ff}

\bibitem[{C.~R. Harris {et~al.}(2020)Harris, Millman, van~der Walt, Gommers,
  Virtanen, Cournapeau, Wieser, Taylor, Berg, Smith, Kern, Picus, Hoyer, van
  Kerkwijk, Brett, Haldane, Fern{\'a}ndez~del R{\'\i}o, Wiebe, Peterson,
  G{\'e}rard-Marchant, Sheppard, Reddy, Weckesser, Abbasi, Gohlke, \&
  Oliphant}]{2020NumPy-Array}
Harris, C.~R., Millman, K.~J., van~der Walt, S.~J., {et~al.} 2020,
  \bibinfo{title}{Array programming with {NumPy},} Nature, 585, 357,
  \dodoi{10.1038/s41586-020-2649-2}

\bibitem[{A.~N. {Heinze} {et~al.}(2015){Heinze}, {Metchev}, \&
  {Trollo}}]{2015AJ....150..125H}
{Heinze}, A.~N., {Metchev}, S., \& {Trollo}, J. 2015, \bibinfo{title}{{Digital
  Tracking Observations Can Discover Asteroids 10 Times Fainter Than
  Conventional Searches},} \aj, 150, 125, \dodoi{10.1088/0004-6256/150/4/125}

\bibitem[{A.~N. {Heinze} {et~al.}(2019){Heinze}, {Trollo}, \&
  {Metchev}}]{2019AJ....158..232H}
{Heinze}, A.~N., {Trollo}, J., \& {Metchev}, S. 2019, \bibinfo{title}{{The Flux
  Distribution and Sky Density of 25th Magnitude Main Belt Asteroids},} \aj,
  158, 232, \dodoi{10.3847/1538-3881/ab48fa}

\bibitem[{M.~J. {Holman} {et~al.}(2004){Holman}, {Kavelaars}, {Grav},
  {Gladman}, {Fraser}, {Milisavljevic}, {Nicholson}, {Burns}, {Carruba},
  {Petit}, {Rousselot}, {Mousis}, {Marsden}, \&
  {Jacobson}}]{2004Natur.430..865H}
{Holman}, M.~J., {Kavelaars}, J.~J., {Grav}, T., {et~al.} 2004,
  \bibinfo{title}{{Discovery of five irregular moons of Neptune},} \nat, 430,
  865, \dodoi{10.1038/nature02832}

\bibitem[{J.~D. Hunter(2007)Hunter}]{Hunter:2007}
Hunter, J.~D. 2007, \bibinfo{title}{Matplotlib: A 2D graphics environment,}
  Computing in Science \& Engineering, 9, 90, \dodoi{10.1109/MCSE.2007.55}

\bibitem[{ {IRSA}(2022){IRSA}}]{https://doi.org/10.26131/irsa539}
{IRSA}. 2022, Zwicky Transient Facility Image Service, IPAC,
  \dodoi{10.26131/IRSA539}

\bibitem[{{\v{Z}}. {Ivezi{\'c}} {et~al.}(2019){Ivezi{\'c}}, {Kahn}, {Tyson},
  {Abel}, {Acosta}, {Allsman}, {Alonso}, {AlSayyad}, {Anderson}, {Andrew},
  {Angel}, {Angeli}, {Ansari}, {Antilogus}, {Araujo}, {Armstrong}, {Arndt},
  {Astier}, {Aubourg}, {Auza}, {Axelrod}, {Bard}, {Barr}, {Barrau}, {Bartlett},
  {Bauer}, {Bauman}, {Baumont}, {Bechtol}, {Bechtol}, {Becker}, {Becla},
  {Beldica}, {Bellavia}, {Bianco}, {Biswas}, {Blanc}, {Blazek}, {Blandford},
  {Bloom}, {Bogart}, {Bond}, {Booth}, {Borgland}, {Borne}, {Bosch}, {Boutigny},
  {Brackett}, {Bradshaw}, {Brandt}, {Brown}, {Bullock}, {Burchat}, {Burke},
  {Cagnoli}, {Calabrese}, {Callahan}, {Callen}, {Carlin}, {Carlson},
  {Chandrasekharan}, {Charles-Emerson}, {Chesley}, {Cheu}, {Chiang}, {Chiang},
  {Chirino}, {Chow}, {Ciardi}, {Claver}, {Cohen-Tanugi}, {Cockrum}, {Coles},
  {Connolly}, {Cook}, {Cooray}, {Covey}, {Cribbs}, {Cui}, {Cutri}, {Daly},
  {Daniel}, {Daruich}, {Daubard}, {Daues}, {Dawson}, {Delgado}, {Dellapenna},
  {de Peyster}, {de Val-Borro}, {Digel}, {Doherty}, {Dubois},
  {Dubois-Felsmann}, {Durech}, {Economou}, {Eifler}, {Eracleous}, {Emmons},
  {Fausti Neto}, {Ferguson}, {Figueroa}, {Fisher-Levine}, {Focke}, {Foss},
  {Frank}, {Freemon}, {Gangler}, {Gawiser}, {Geary}, {Gee}, {Geha}, {Gessner},
  {Gibson}, {Gilmore}, {Glanzman}, {Glick}, {Goldina}, {Goldstein}, {Goodenow},
  {Graham}, {Gressler}, {Gris}, {Guy}, {Guyonnet}, {Haller}, {Harris},
  {Hascall}, {Haupt}, {Hernandez}, {Herrmann}, {Hileman}, {Hoblitt}, {Hodgson},
  {Hogan}, {Howard}, {Huang}, {Huffer}, {Ingraham}, {Innes}, {Jacoby}, {Jain},
  {Jammes}, {Jee}, {Jenness}, {Jernigan}, {Jevremovi{\'c}}, {Johns}, {Johnson},
  {Johnson}, {Jones}, {Juramy-Gilles}, {Juri{\'c}}, {Kalirai}, {Kallivayalil},
  {Kalmbach}, {Kantor}, {Karst}, {Kasliwal}, {Kelly}, {Kessler}, {Kinnison},
  {Kirkby}, {Knox}, {Kotov}, {Krabbendam}, {Krughoff}, {Kub{\'a}nek},
  {Kuczewski}, {Kulkarni}, {Ku}, {Kurita}, {Lage}, {Lambert}, {Lange},
  {Langton}, {Le Guillou}, {Levine}, {Liang}, {Lim}, {Lintott}, {Long},
  {Lopez}, {Lotz}, {Lupton}, {Lust}, {MacArthur}, {Mahabal}, {Mandelbaum},
  {Markiewicz}, {Marsh}, {Marshall}, {Marshall}, {May}, {McKercher}, {McQueen},
  {Meyers}, {Migliore}, {Miller}, \& {Mills}}]{2019ApJ...873..111I}
{Ivezi{\'c}}, {\v{Z}}., {Kahn}, S.~M., {Tyson}, J.~A., {et~al.} 2019,
  \bibinfo{title}{{LSST: From Science Drivers to Reference Design and
  Anticipated Data Products},} \apj, 873, 111, \dodoi{10.3847/1538-4357/ab042c}

\bibitem[{R.~L. {Jones} {et~al.}(2018){Jones}, {Slater}, {Moeyens}, {Allen},
  {Axelrod}, {Cook}, {Ivezi{\'c}}, {Juri{\'c}}, {Myers}, \&
  {Petry}}]{2018Icar..303..181J}
{Jones}, R.~L., {Slater}, C.~T., {Moeyens}, J., {et~al.} 2018,
  \bibinfo{title}{{The Large Synoptic Survey Telescope as a Near-Earth Object
  discovery machine},} \icarus, 303, 181, \dodoi{10.1016/j.icarus.2017.11.033}

\bibitem[{N. {Kaiser} {et~al.}(2002){Kaiser}, {Aussel}, {Burke}, {Boesgaard},
  {Chambers}, {Chun}, {Heasley}, {Hodapp}, {Hunt}, {Jedicke}, {Jewitt},
  {Kudritzki}, {Luppino}, {Maberry}, {Magnier}, {Monet}, {Onaka}, {Pickles},
  {Rhoads}, {Simon}, {Szalay}, {Szapudi}, {Tholen}, {Tonry}, {Waterson}, \&
  {Wick}}]{2002SPIE.4836..154K}
{Kaiser}, N., {Aussel}, H., {Burke}, B.~E., {et~al.} 2002,
  \bibinfo{title}{{Pan-STARRS: A Large Synoptic Survey Telescope Array},} in
  Society of Photo-Optical Instrumentation Engineers (SPIE) Conference Series,
  Vol. 4836, Survey and Other Telescope Technologies and Discoveries, ed. J.~A.
  {Tyson} \& S.~{Wolff}, 154--164, \dodoi{10.1117/12.457365}

\bibitem[{I.~R. {King}(1983){King}}]{1983PASP...95..163K}
{King}, I.~R. 1983, \bibinfo{title}{{Accuracy of measurement of star images on
  a pixel array.},} \pasp, 95, 163, \dodoi{10.1086/131139}

\bibitem[{D. {Lang} {et~al.}(2009){Lang}, {Hogg}, {Jester}, \&
  {Rix}}]{2009AJ....137.4400L}
{Lang}, D., {Hogg}, D.~W., {Jester}, S., \& {Rix}, H.-W. 2009,
  \bibinfo{title}{{Measuring the Undetectable: Proper Motions and Parallaxes of
  Very Faint Sources},} \aj, 137, 4400, \dodoi{10.1088/0004-6256/137/5/4400}

\bibitem[{S. {Li} {et~al.}(2025){Li} {et~al.}}]{sage2025inprep}
{Li}, S., {et~al.} 2025, In preparation

\bibitem[{N. {Lifset} {et~al.}(2021){Lifset}, {Golovich}, {Green}, {Armstrong},
  \& {Yeager}}]{2021AJ....161..282L}
{Lifset}, N., {Golovich}, N., {Green}, E., {Armstrong}, R., \& {Yeager}, T.
  2021, \bibinfo{title}{{A Search for L4 Earth Trojan Asteroids Using a Novel
  Track-before-detect Multiepoch Pipeline},} \aj, 161, 282,
  \dodoi{10.3847/1538-3881/abf7af}

\bibitem[{F.~J. {Masci} {et~al.}(2019){Masci}, {Laher}, {Rusholme}, {Shupe},
  {Groom}, {Surace}, {Jackson}, {Monkewitz}, {Beck}, {Flynn}, {Terek},
  {Landry}, {Hacopians}, {Desai}, {Howell}, {Brooke}, {Imel}, {Wachter}, {Ye},
  {Lin}, {Cenko}, {Cunningham}, {Rebbapragada}, {Bue}, {Miller}, {Mahabal},
  {Bellm}, {Patterson}, {Juri{\'c}}, {Golkhou}, {Ofek}, {Walters}, {Graham},
  {Kasliwal}, {Dekany}, {Kupfer}, {Burdge}, {Cannella}, {Barlow}, {Van
  Sistine}, {Giomi}, {Fremling}, {Blagorodnova}, {Levitan}, {Riddle}, {Smith},
  {Helou}, {Prince}, \& {Kulkarni}}]{2019PASP..131a8003M}
{Masci}, F.~J., {Laher}, R.~R., {Rusholme}, B., {et~al.} 2019,
  \bibinfo{title}{{The Zwicky Transient Facility: Data Processing, Products,
  and Archive},} \pasp, 131, 018003, \dodoi{10.1088/1538-3873/aae8ac}

\bibitem[{P. {McGill} {et~al.}(2025){McGill} {et~al.}}]{mcgill2025inprep}
{McGill}, P., {et~al.} 2025, In preparation

\bibitem[{J.~E. Meyers {et~al.}(2025)Meyers, Schneider, Ebert, Schlafly,
  Yeager, Perloff, Merl, Lifset, Bernstein, Dawson, Golovich, Higgins, McGill,
  Miller, \& Pruett}]{ssapy1}
Meyers, J.~E., Schneider, M.~D., Ebert, J.~T., {et~al.} 2025,
  \bibinfo{title}{SSAPy - Space Situational Awareness for Python,} Journal of
  Open Source Software, 10, 8147, \dodoi{10.21105/joss.08147}

\bibitem[{A. {Milani}(1999){Milani}}]{1999Icar..137..269M}
{Milani}, A. 1999, \bibinfo{title}{{The Asteroid Identification Problem. I.
  Recovery of Lost Asteroids},} \icarus, 137, 269,
  \dodoi{10.1006/icar.1999.6045}

\bibitem[{A. {Milani} {et~al.}(2005){Milani}, {Sansaturio}, {Tommei},
  {Arratia}, \& {Chesley}}]{2005A&A...431..729M}
{Milani}, A., {Sansaturio}, M.~E., {Tommei}, G., {Arratia}, O., \& {Chesley},
  S.~R. 2005, \bibinfo{title}{{Multiple solutions for asteroid orbits:
  Computational procedure and applications},} \aap, 431, 729,
  \dodoi{10.1051/0004-6361:20041737}

\bibitem[{K. {Muinonen}(1996){Muinonen}}]{1996MNRAS.280.1235M}
{Muinonen}, K. 1996, \bibinfo{title}{{Orbital covariance eigenproblem for
  asteroids and comets},} \mnras, 280, 1235, \dodoi{10.1093/mnras/280.4.1235}

\bibitem[{K. {Muinonen} \& E. {Bowell}(1993){Muinonen} \&
  {Bowell}}]{1993Icar..104..255M}
{Muinonen}, K., \& {Bowell}, E. 1993, \bibinfo{title}{{Asteroid Orbit
  Determination Using Bayesian Probabilities},} \icarus, 104, 255,
  \dodoi{10.1006/icar.1993.1100}

\bibitem[{K. {Muinonen} {et~al.}(2006){Muinonen}, {Virtanen}, {Granvik}, \&
  {Laakso}}]{2006MNRAS.368..809M}
{Muinonen}, K., {Virtanen}, J., {Granvik}, M., \& {Laakso}, T. 2006,
  \bibinfo{title}{{Asteroid orbits using phase-space volumes of variation},}
  \mnras, 368, 809, \dodoi{10.1111/j.1365-2966.2006.10168.x}

\bibitem[{B.~J. {Owen}(1996){Owen}}]{1996PhRvD..53.6749O}
{Owen}, B.~J. 1996, \bibinfo{title}{{Search templates for gravitational waves
  from inspiraling binaries: Choice of template spacing},} \prd, 53, 6749,
  \dodoi{10.1103/PhysRevD.53.6749}

\bibitem[{A.~H. {Parker} \& J.~J. {Kavelaars}(2010){Parker} \&
  {Kavelaars}}]{2010PASP..122..549P}
{Parker}, A.~H., \& {Kavelaars}, J.~J. 2010, \bibinfo{title}{{Pencil-Beam
  Surveys for Trans-Neptunian Objects: Novel Methods for Optimization and
  Characterization},} \pasp, 122, 549, \dodoi{10.1086/652424}

\bibitem[{W.~H. Press {et~al.}(2007)Press, Teukolsky, Vetterling, \&
  Flannery}]{Press2007}
Press, W.~H., Teukolsky, S.~A., Vetterling, W.~T., \& Flannery, B.~P. 2007,
  Numerical Recipes 3rd Edition: The Art of Scientific Computing, 3rd edn.
  (Cambridge University Press)

\bibitem[{D.~L. {Rabinowitz} {et~al.}(2007){Rabinowitz}, {Schaefer}, \&
  {Tourtellotte}}]{2007AJ....133...26R}
{Rabinowitz}, D.~L., {Schaefer}, B.~E., \& {Tourtellotte}, S.~W. 2007,
  \bibinfo{title}{{The Diverse Solar Phase Curves of Distant Icy Bodies. I.
  Photometric Observations of 18 Trans-Neptunian Objects, 7 Centaurs, and
  Nereid},} \aj, 133, 26, \dodoi{10.1086/508931}

\bibitem[{M. {Rice} \& G. {Laughlin}(2020){Rice} \&
  {Laughlin}}]{2020PSJ.....1...81R}
{Rice}, M., \& {Laughlin}, G. 2020, \bibinfo{title}{{Exploring Trans-Neptunian
  Space with TESS: A Targeted Shift-stacking Search for Planet Nine and Distant
  TNOs in the Galactic Plane},} \psj, 1, 81, \dodoi{10.3847/PSJ/abc42c}

\bibitem[{G.~B. {Rybicki} \& W.~H. {Press}(1992){Rybicki} \&
  {Press}}]{1992ApJ...398..169R}
{Rybicki}, G.~B., \& {Press}, W.~H. 1992, \bibinfo{title}{{Interpolation,
  Realization, and Reconstruction of Noisy, Irregularly Sampled Data},} \apj,
  398, 169, \dodoi{10.1086/171845}

\bibitem[{B.~S. {Sathyaprakash} \& S.~V. {Dhurandhar}(1991){Sathyaprakash} \&
  {Dhurandhar}}]{1991PhRvD..44.3819S}
{Sathyaprakash}, B.~S., \& {Dhurandhar}, S.~V. 1991, \bibinfo{title}{{Choice of
  filters for the detection of gravitational waves from coalescing binaries},}
  \prd, 44, 3819, \dodoi{10.1103/PhysRevD.44.3819}

\bibitem[{C. {Shankman} {et~al.}(2017){Shankman}, {Kavelaars}, {Bannister},
  {Gladman}, {Lawler}, {Chen}, {Jakubik}, {Kaib}, {Alexandersen}, {Gwyn},
  {Petit}, \& {Volk}}]{2017AJ....154...50S}
{Shankman}, C., {Kavelaars}, J.~J., {Bannister}, M.~T., {et~al.} 2017,
  \bibinfo{title}{{OSSOS. VI. Striking Biases in the Detection of Large
  Semimajor Axis Trans-Neptunian Objects},} \aj, 154, 50,
  \dodoi{10.3847/1538-3881/aa7aed}

\bibitem[{M. {Shao} {et~al.}(2014){Shao}, {Nemati}, {Zhai}, {Turyshev},
  {Sandhu}, {Hallinan}, \& {Harding}}]{2014ApJ...782....1S}
{Shao}, M., {Nemati}, B., {Zhai}, C., {et~al.} 2014, \bibinfo{title}{{Finding
  Very Small Near-Earth Asteroids using Synthetic Tracking},} \apj, 782, 1,
  \dodoi{10.1088/0004-637X/782/1/1}

\bibitem[{A. {Siraj} {et~al.}(2025){Siraj}, {Chyba}, \&
  {Tremaine}}]{2025ApJ...978..139S}
{Siraj}, A., {Chyba}, C.~F., \& {Tremaine}, S. 2025, \bibinfo{title}{{Orbit of
  a Possible Planet X},} \apj, 978, 139, \dodoi{10.3847/1538-4357/ad98f6}

\bibitem[{K.~M. {Smith}(2016){Smith}}]{2016arXiv161006831S}
{Smith}, K.~M. 2016, \bibinfo{title}{{New algorithms for radio pulsar search},}
  arXiv e-prints, arXiv:1610.06831, \dodoi{10.48550/arXiv.1610.06831}

\bibitem[{H. {Smotherman} {et~al.}(2024){Smotherman}, {Bernardinelli},
  {Portillo}, {Connolly}, {Kalmbach}, {Stetzler}, {Juri{\'c}},
  {Bekte{\v{s}}evi{\'c}}, {Langford}, {Adams}, {Oldroyd}, {Holman}, {Chandler},
  {Fuentes}, {Gerdes}, {Lin}, {Markwardt}, {McNeill}, {Mommert}, {Napier},
  {Payne}, {Ragozzine}, {Rivkin}, {Schlichting}, {Sheppard}, {Strauss},
  {Trilling}, \& {Trujillo}}]{2024AJ....167..136S}
{Smotherman}, H., {Bernardinelli}, P.~H., {Portillo}, S. K.~N., {et~al.} 2024,
  \bibinfo{title}{{The DECam Ecliptic Exploration Project (DEEP). VI. First
  Multiyear Observations of Trans-Neptunian Objects},} \aj, 167, 136,
  \dodoi{10.3847/1538-3881/ad1524}

\bibitem[{ {The pandas development team}(2020){The pandas development
  team}}]{reback2020pandas}
{The pandas development team}. 2020, pandas-dev/pandas: Pandas, 2.2.1 Zenodo,
  \dodoi{10.5281/zenodo.3509134}

\bibitem[{J.~L. {Tonry} {et~al.}(2012){Tonry}, {Stubbs}, {Lykke}, {Doherty},
  {Shivvers}, {Burgett}, {Chambers}, {Hodapp}, {Kaiser}, {Kudritzki},
  {Magnier}, {Morgan}, {Price}, \& {Wainscoat}}]{2012ApJ...750...99T}
{Tonry}, J.~L., {Stubbs}, C.~W., {Lykke}, K.~R., {et~al.} 2012,
  \bibinfo{title}{{The Pan-STARRS1 Photometric System},} \apj, 750, 99,
  \dodoi{10.1088/0004-637X/750/2/99}

\bibitem[{P. Virtanen {et~al.}(2020)Virtanen, Gommers, Oliphant, Haberland,
  Reddy, Cournapeau, Burovski, Peterson, Weckesser, Bright, {van der Walt},
  Brett, Wilson, Millman, Mayorov, Nelson, Jones, Kern, Larson, Carey, Polat,
  Feng, Moore, {VanderPlas}, Laxalde, Perktold, Cimrman, Henriksen, Quintero,
  Harris, Archibald, Ribeiro, Pedregosa, {van Mulbregt}, \& {SciPy 1.0
  Contributors}}]{2020SciPy-NMeth}
Virtanen, P., Gommers, R., Oliphant, T.~E., {et~al.} 2020,
  \bibinfo{title}{{{SciPy} 1.0: Fundamental Algorithms for Scientific Computing
  in Python},} Nature Methods, 17, 261, \dodoi{10.1038/s41592-019-0686-2}

\bibitem[{B. {Wang} {et~al.}(2018){Wang}, {Zhao}, \&
  {Li}}]{2018ChA&A..42..433W}
{Wang}, B., {Zhao}, H.-b., \& {Li}, B. 2018, \bibinfo{title}{{Detection of
  Faint Asteroids Based on Image Shifting and Stacking Method},} \caa, 42, 433,
  \dodoi{10.1016/j.chinastron.2018.09.007}

\bibitem[{S. {Weinberg}(1972){Weinberg}}]{1972gcpa.book.....W}
{Weinberg}, S. 1972, {Gravitation and Cosmology: Principles and Applications of
  the General Theory of Relativity} ({John Wiley \& Sons, Inc.})

\bibitem[{ {W}es {M}c{K}inney(2010){W}es
  {M}c{K}inney}]{mckinney-proc-scipy-2010}
{W}es {M}c{K}inney. 2010, \bibinfo{title}{{D}ata {S}tructures for {S}tatistical
  {C}omputing in {P}ython,} in {P}roceedings of the 9th {P}ython in {S}cience
  {C}onference, ed. {S}t\'efan van~der {W}alt \& {J}arrod {M}illman, 56 -- 61,
  \dodoi{10.25080/Majora-92bf1922-00a}

\bibitem[{P.~J. {Whidden} {et~al.}(2019){Whidden}, {Bryce Kalmbach},
  {Connolly}, {Jones}, {Smotherman}, {Bektesevic}, {Slater}, {Becker},
  {Ivezi{\'c}}, {Juri{\'c}}, {Bolin}, {Moeyens}, {F{\"o}rster}, \&
  {Golkhou}}]{2019AJ....157..119W}
{Whidden}, P.~J., {Bryce Kalmbach}, J., {Connolly}, A.~J., {et~al.} 2019,
  \bibinfo{title}{{Fast Algorithms for Slow Moving Asteroids: Constraints on
  the Distribution of Kuiper Belt Objects},} \aj, 157, 119,
  \dodoi{10.3847/1538-3881/aafd2d}

\bibitem[{C. {Zhai} {et~al.}(2014){Zhai}, {Shao}, {Nemati}, {Werne}, {Zhou},
  {Turyshev}, {Sandhu}, {Hallinan}, \& {Harding}}]{2014ApJ...792...60Z}
{Zhai}, C., {Shao}, M., {Nemati}, B., {et~al.} 2014, \bibinfo{title}{{Detection
  of a Faint Fast-moving Near-Earth Asteroid Using the Synthetic Tracking
  Technique},} \apj, 792, 60, \dodoi{10.1088/0004-637X/792/1/60}

\bibitem[{C. {Zhai} {et~al.}(2018){Zhai}, {Shao}, {Saini}, {Sandhu}, {Owen},
  {Choi}, {Werne}, {Ely}, {Lazio}, {Martin-Mur}, {Preston}, {Turyshev},
  {Mitchell}, {Nazli}, {Cui}, \& {Mochama}}]{2018AJ....156...65Z}
{Zhai}, C., {Shao}, M., {Saini}, N.~S., {et~al.} 2018,
  \bibinfo{title}{{Accurate Ground-based Near-Earth-Asteroid Astrometry Using
  Synthetic Tracking},} \aj, 156, 65, \dodoi{10.3847/1538-3881/aacb28}

\bibitem[{C. {Zhai} {et~al.}(2020){Zhai}, {Ye}, {Shao}, {Trahan}, {Saini},
  {Shen}, {Prince}, {Bellm}, {Graham}, {Helou}, {Kulkarni}, {Kupfer}, {Laher},
  {Mahabal}, {Masci}, {Rusholme}, {Rosnet}, \& {Shupe}}]{2020PASP..132f4502Z}
{Zhai}, C., {Ye}, Q., {Shao}, M., {et~al.} 2020, \bibinfo{title}{{Synthetic
  Tracking Using ZTF Deep Drilling Data Sets},} \pasp, 132, 064502,
  \dodoi{10.1088/1538-3873/ab828b}

\end{thebibliography}
\bibliographystyle{aasjournalv7}



\end{document}